\documentclass[fleqn,usenatbib]{mnras}

\usepackage{newtxtext,newtxmath}

\usepackage[T1]{fontenc}

\DeclareRobustCommand{\VAN}[3]{#2}
\let\VANthebibliography\thebibliography
\def\thebibliography{\DeclareRobustCommand{\VAN}[3]{##3}\VANthebibliography}

\usepackage{graphicx}	
\usepackage{amsmath}	
\usepackage{multirow}
\usepackage{multicol}
\usepackage{bigstrut}
\usepackage[normal]{threeparttable}

\newcommand{\hydra}{\textsc{h}{\small y}\textsc{dra} }
\newcommand{\hydrar}{\textsc{h}{\small y}\textsc{dr}{\small o} }

\title[HyDRo: Atmospheric Retrieval of Rocky Exoplanets]{HyDRo: Atmospheric Retrieval of Rocky Exoplanets in Thermal Emission}

\author[A. A. A. Piette et al.]{
Anjali A. A. Piette,$^{1}$\thanks{E-mail: ap763@cam.ac.uk}
Nikku Madhusudhan,$^{1}$\thanks{E-mail: nmadhu@ast.cam.ac.uk}
Avi M. Mandell$^{2}$
\\
$^{1}$Institute of Astronomy, University of Cambridge, Madingley Road, Cambridge, CB3 0HA, UK\\
$^{2}$Solar System Exploration Division, NASA Goddard Space Flight Center, Greenbelt, MD 20771, USA\\
}

\date{Accepted XXX. Received YYY; in original form ZZZ}

\pubyear{2021}

\begin{document}
\label{firstpage}
\pagerange{\pageref{firstpage}--\pageref{lastpage}}
\maketitle

\begin{abstract}
Emission spectroscopy is a promising technique to observe atmospheres of rocky exoplanets, probing both their chemistry and thermal profiles. We present \textsc{h}{\small y}\textsc{dr}{\small o}, an atmospheric retrieval framework for thermal emission spectra of rocky exoplanets. \hydrar does not make prior assumptions about the background atmospheric composition, and can therefore be used to interpret spectra of secondary atmospheres with unknown compositions. We use \hydrar to assess the chemical constraints which can be placed on rocky exoplanet atmospheres using JWST. Firstly, we identify the best currently-known rocky exoplanet candidates for spectroscopic observations in thermal emission with JWST, finding $>30$ known rocky exoplanets whose thermal emission will be detectable by JWST/MIRI in fewer than 10 eclipses at $R\sim10$. We then consider the observations required to characterise the atmospheres of three promising rocky exoplanets across the $\sim$400-800~K equilibrium temperature range: Trappist-1~b, GJ~1132~b, and LHS~3844~b. Considering a range of CO$_2$- to H$_2$O-rich atmospheric compositions, we find that as few as 8 eclipses of LHS~3844~b or GJ~1132~b with MIRI will be able to place important constraints on the chemical compositions of their atmospheres. This includes confident detections of CO$_2$ and H$_2$O in the case of a cloud-free CO$_2$-rich composition, besides ruling out a bare rock scenario. Similarly, 30 eclipses of Trappist-1~b with MIRI/LRS can allow detections of a cloud-free CO$_2$-rich or CO$_2$-H$_2$O atmosphere. \hydrar will allow important atmospheric constraints for rocky exoplanets using JWST observations, providing clues about their geochemical environments.
\end{abstract}




\section{Introduction}
\label{sec:intro}

In recent years, the atmospheric characterisation of exoplanets has flourished, with increasingly detailed constraints made possible by unprecedented observations and sophisticated atmospheric retrieval tools and models \citep[e.g.][]{Crossfield2015,Kreidberg2018haex,Madhusudhan2019}. With the upcoming James Webb Space Telescope (JWST), the next frontier of atmospheric characterisation will turn towards smaller and cooler planets, including rocky exoplanets potentially hosting terrestrial-like secondary atmospheres \citep[e.g.][]{Greene2016,Lustig-Yaeger2019,Turbet2020}. In this work, we report a new retrieval framework designed to analyse the thermal emission spectra of rocky exoplanets. We further use this framework to assess the ideal targets and observations needed to make important new constraints on rocky exoplanet atmospheric chemistry. 

Recent observations have probed the atmospheres of several rocky exoplanets. For example, transmission spectroscopy of GJ~1132~b has suggested an atmosphere with a  high-mean-molecular-weight (high-$\mu$) and/or high-altitude clouds \citep{Southworth2017,Diamond-Lowe2018,Mugnai2021}. Similarly, transmission spectra of Trappist-1~d, e and f have ruled out clear, H$_2$-rich atmospheres for these planets \citep{Dewit2018,Moran2018}. Furthermore, searches for molecular species such as HCN have now become possible in the atmospheres of rocky exoplanets such as 55~Cnc~e and GJ~1132~b \citep{Tsiaras2016,Swain2021,Mugnai2021} and could provide insights into their possible interior and atmospheric conditions \citep[e.g.][]{Madhusudhan2012_55cnce}.  

Thermal phase curves have also been used to place constraints on rocky exoplanet atmospheres. The phase curve of 55~Cnc~e displays a strong hot-spot shift, indicating the presence of an atmosphere \citep{Demory2016,Hammond2017,Angelo2017}. In contrast, the strong day-night contrast and lack of hot-spot shift in the phase curve of LHS~3844~b suggests that this planet does not host a clear, H$_2$-rich atmosphere deeper than $\sim$0.1~bar and may instead host a high-mean-molecular-weight and/or cloudy atmosphere, or no atmosphere at all \citep{Kreidberg2019,Diamond-Lowe2020}. Future observations, e.g. with JWST, will provide chemical constraints on rocky exoplanet atmospheres, leading to unprecedented constraints on their formation and evolution. 

Thermal emission spectroscopy in particular offers an exceptional opportunity to characterise secondary, high-$\mu$ atmospheres on rocky exoplanets. This method is highly sensitive to atmospheric temperature profiles as well as molecular absorption. The mid-infrared wavelength range is particularly advantageous in the study of secondary atmospheres given the abundance of molecular features in this range \citep[e.g.][]{Deming2009a,Greene2016,Madhusudhan2019}. Furthermore, the signal-to-noise ratio (S/N) achievable for secondary eclipse observations is maximised in the mid-infrared. JWST's Mid-Infrared Instrument (MIRI), with a spectral range of $\sim$5-28$\mu$m, will therefore provide an unprecedented opportunity to investigate rocky exoplanet atmospheres including their chemical and thermal conditions.

Theoretical models of rocky planet atmospheres and surface-atmosphere interactions predict a wide range of possible atmospheric compositions beyond those seen in the solar system \citep{Leconte2015}. For example, surface-atmosphere interactions could result in atmospheres composed of H$_2$O, CO$_2$, SO$_2$, SiO, O$_2$ and H$_2$ depending on the surface conditions and bulk planetary properties \citep[e.g.][]{Gaillard2014,Dorn2018a,Herbort2020,Lichtenberg2021,Schlichting2021}. Degassing during accretion can further result in a wide range of atmospheric compositions including hydrogen, water and carbon compounds \citep{Elkins-Tanton2008}. Furthermore, \citet{Hu2012} find that UV irradiation from an M-dwarf such as Trappist-1 could drive chemical reactions producing $>$1~bar of CO and/or O$_2$. The interpretation of rocky exoplanet atmospheric observations therefore requires an approach which encompasses this chemical diversity.

Atmospheric retrievals have revolutionised atmospheric characterisation of exoplanets in the past decade \citep{Madhusudhan2018} and will be critical to provide detailed constraints on rocky exoplanet atmospheres. To date, retrievals of exoplanet emission spectra have typically been applied to hydrogen-rich atmospheres \citep[e.g.][]{Madhusudhan2009,Madhusudhan2010,
Lee2012,Line2013,Waldmann2015,Lavie2017,Gandhi2018,Molliere2019} and do not consider secondary atmospheric compositions dominated by non-H species. However, \citet{Benneke2012} report an agnostic chemical abundance parameterisation which does not assume a dominant atmospheric component a priori, and has been applied in the context of transmission spectroscopy \citep{Benneke2012,Benneke2013,Welbanks2021}. By using the centred-log-ratio transformation, the parameterisation applies an identical prior probability distribution to each chemical species in the retrieval, allowing it to constrain the compositions of secondary atmospheres.

In this work, we present the first atmospheric retrieval framework for the thermal emission spectra of rocky exoplanets. The framework utilises the \hydra emission retrieval framework of \citet{Gandhi2018} and the chemical abundance parameterisation of \citet{Benneke2012}. It includes a range of opacity sources expected in secondary atmospheres, for example allowing for Venus-like and H$_2$O-dominated compositions. We further identify a large sample of optimal rocky exoplanet candidates for atmospheric characterisation in thermal emission with JWST MIRI. Focusing on three case studies, we use the new retrieval framework to investigate the chemical detections which can be made using JWST MIRI. In particular, we consider a hierarchy of science cases which can be used to systematically confirm/exclude increasingly complex atmospheric states.

In what follows, we describe the new retrieval framework in Section \ref{sec:methods_ret}. In Section \ref{sec:methods_spectra}, we outline the self-consistent atmospheric model used to model rocky exoplanet thermal emission spectra, and use such a model in section \ref{sec:validation} to validate the retrieval framework. We identify optimal rocky exoplanet candidates for atmospheric characterisation in Section \ref{sec:targets}. In Section \ref{sec:case_studies}, we investigate the observability of CO$_2$ and H$_2$O in three of the best targets for atmospheric characterization with MIRI in different temperature regimes: LHS~3844~b, GJ~1132~b and Trappist-1~b. We discuss further considerations in Section \ref{sec:discussion} and present our conclusions in Section \ref{sec:conclusions}. 


\section{Retrieval Framework \& Atmospheric Model}
\label{sec:model}
We describe here the \hydrar retrieval framework, adapted for use with rocky planets whose primary atmospheric constituent is unknown (Section \ref{sec:methods_ret}). To test this framework and to provide predictions and observability strategies for JWST, we calculate self-consistent atmospheric models for known rocky exoplanets, and simulate their observed thermal emission spectra with JWST in Sections \ref{sec:validation} and \ref{sec:case_studies}. These atmospheric models and simulated data are described in Section \ref{sec:methods_spectra}.

\subsection{Retrieval Framework}
\label{sec:methods_ret}
In this work, we build on the \hydra retrieval framework \citep{Gandhi2018} and adapt it for use with rocky exoplanets. \hydra includes a parametric 1-D forward model coupled to a Nested Sampling Bayesian parameter estimation algorithm \citep{Skilling2006}, \textsc{PyMultiNest} \citep{Feroz2009,Buchner2014}. The inputs to the parametric forward model typically consist of a model pressure-temperature ($P$-$T$) profile and chemical abundances, which are used to calculate the emergent thermal emission spectrum of the planet. As in \citet{Gandhi2018}, we use the 6-parameter $P$-$T$ profile of \citet{Madhusudhan2009}, which we find to be successful in retrieving the self-consistently derived temperature profiles of rocky exoplanets (Sections \ref{sec:validation} and \ref{sec:case_studies}). The priors we use for each of the $P$-$T$ profile parameters are shown in Table \ref{tab:PTpriors}. These priors allow for both inverted and non-inverted temperature profiles in order to span the wide range of temperature gradients which can exist in the dayside atmosphere. As well as providing posterior probability distributions for the model parameters, the Nested Sampling algorithm further calculates the Bayesian evidence of the model fit. This can be used to perform robust model comparisons and to assess the statistical significance of molecular detections.

To date, \hydra and adaptations thereof have been used to retrieve the thermal emission spectra of H$_2$-dominated atmospheres, including hot Jupiters and brown dwarfs \citep{Gandhi2018, Gandhi2019b, Piette2020b}. For these cases, the retrieval implicitly assumes that the background molecule is H$_2$. In the case of rocky exoplanet atmospheres, however, the primary constituent of the atmosphere is unknown and a background molecule cannot be assumed. This motivates several adaptations to the retrieval framework, which are described in Sections \ref{sec:ret_chemparam} and \ref{sec:ret_opac} below. Unlike giant planets and brown dwarfs, rocky planets also have the potential to host little or no atmosphere at all due to atmospheric escape processes \citep[e.g.][]{Kreidberg2019}. In Section \ref{sec:ret_detect}, we discuss how the Bayesian evidences of rocky exoplanet retrievals can be used to statistically assess the presence of an atmosphere, as well as to calculate the detection significance for specific chemical species. We further validate the \hydrar retrieval framework in section \ref{sec:validation} by applying it to simulated data for Trappist-1~b.

\begin{table}
    \centering
    \caption{Prior probability distributions for the $P$-$T$ profile parameters (see also \citealt{Madhusudhan2009}, \citealt{Gandhi2018}).}
    \begin{tabular}{c|c|c}
       \hline
       \bf Parameter  &  \bf Prior Distribution & \bf Range\\
       \hline
       \hline
       $\alpha_1$/K$^{-1/2}$ & uniform & 0.02 $-$ 1 \\
       $\alpha_2$/K$^{-1/2}$ & uniform & 0.02 $-$ 1\\
       $T_{100 \rm mb}$/K & uniform & 100 $-$ 2500\\
       $P_1$/bar & log-uniform & $10^{-5}$ $-$ 100\\
       $P_2$/bar & log-uniform & $10^{-5}$ $-$ 100\\
       $P_3$/bar & log-uniform & $10^{-2}$ $-$ 100\\
       \hline
    \end{tabular}
    \label{tab:PTpriors}
\end{table}

\subsubsection{Chemical abundance parameterisation}
\label{sec:ret_chemparam}

\begin{figure*}
    \centering
    \includegraphics[width=0.8\textwidth]{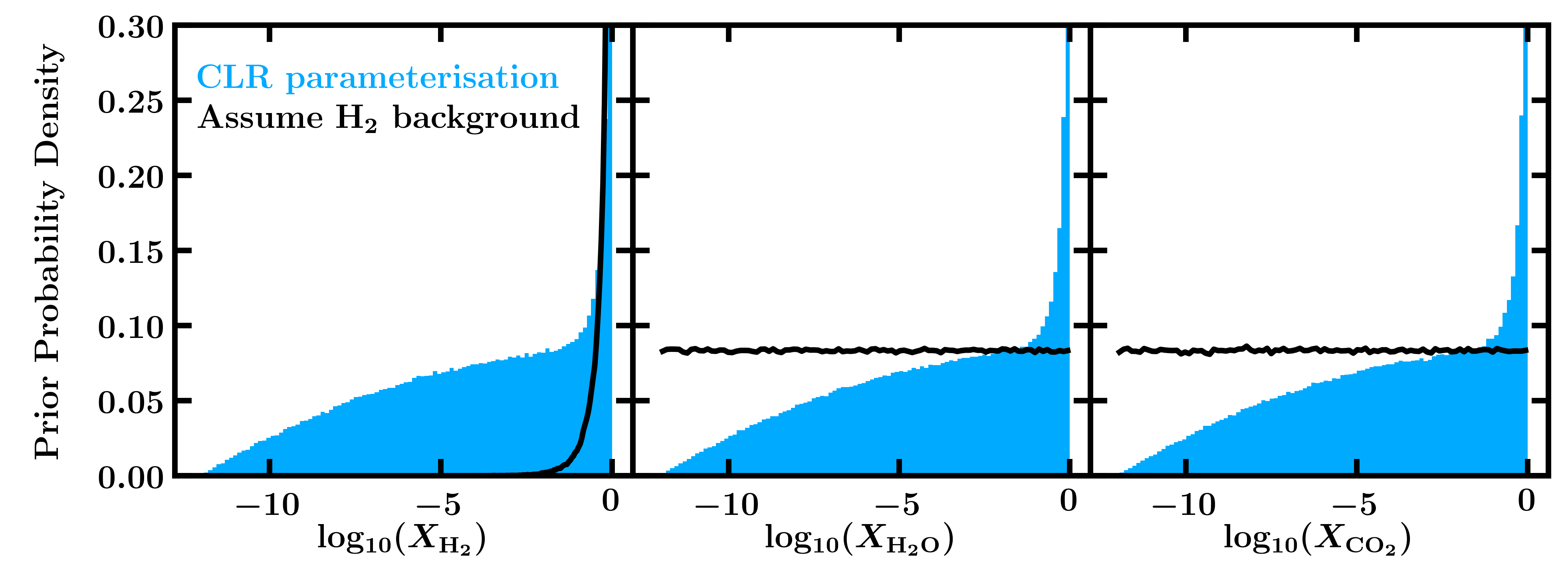}
    \caption{Prior probability densities for log chemical abundances using the centred-log-ratio (CLR) parameterisation (blue) or assuming a H$_2$-dominated composition and log-normal priors for the other species (black). The CLR parameterisation results in identical priors for all species, without assuming a fixed background species.}
    \label{fig:priors}
\end{figure*}

In retrievals of H$_2$-dominated atmospheres, the abundances of trace gases are typically included as parameters with a log-uniform prior. The mixing ratio of H$_2$, $X_\mathrm{H_2}$, is then calculated given the constraint that all mixing ratios, $X_i$, must sum to one, i.e. 
\begin{equation}
    \label{eq:sum1}
    \sum_{i=1}^{n} X_i = 1
\end{equation}
where $n$ is the number of chemical species in the model. This treatment results in a different prior for the mixing ratio of the background molecule (in this case $X_\mathrm{H_2}$) relative to the mixing ratios of the other species. This is shown in Figure \ref{fig:priors}; in particular, the prior for $X_\mathrm{H_2}$ is heavily skewed to higher values relative to the other chemical species. While appropriate for giant planets and brown dwarfs, these priors would result in biased results for low-mass exoplanets whose primary atmospheric constituent is unknown.

\citet{Benneke2012} proposed an alternative parameterisation for the chemical mixing ratios in exoplanets with unknown dominant atmospheric constituents, and applied it to retrievals of transmission spectra of super-Earths. This parameterisation avoids the issues described above by using the centred-log-ratio (CLR) transformation, which is commonly used in the geological sciences to interpret compositional data \citep[e.g.][]{Pawlowsky-Glahn2006}. Here, we adopt this method for retrievals of thermal emission spectra. As described by \citet{Benneke2012}, the transformed `CLR' parameters for each of the $n$ chemical species included in the model, $\xi_i$, are given by
\begin{equation}
    \label{eq:xi_i}
    \xi_i = \ln \frac{X_i}{g(\bf x)},
\end{equation}
where
\begin{equation}
    \label{eq:g_x}
    g({\bf x}) = \exp \left(\frac{1}{n} \sum_{i=1}^{n} \ln X_i \right).
\end{equation}
The $\xi_i$ parameters are then sampled by the Bayesian parameter estimation algorithm assuming uniform priors. In principle, $\xi_i$ can vary between $-\infty$ and $+\infty$, corresponding to mixing ratios between 0 and 1. However, in practice, the chemical species have vanishing effects on the model spectrum once very small abundances are reached. For a given assumed minimum mixing ratio, $p_\mathrm{min}$, the corresponding maximum possible mixing ratio for a given species is $1-p_\mathrm{min}(n-1)$, following equation \ref{eq:sum1}. For the $\xi_i$ parameters, these limits translate to 
\begin{equation*}
    \frac{1}{n}\left[\ln(p_\mathrm{min})-\ln(1-np_\mathrm{min})\right]
\end{equation*}
for the lower bound of the uniform prior, and
\begin{equation*}
    \frac{n-1}{n}\left[\ln(1-np_\mathrm{min})-\ln(p_\mathrm{min})\right]
\end{equation*}
for the upper bound. In this work, we choose to use $p_\mathrm{min} = 10^{-12}$, as in \citet{Benneke2012}. Once the $\xi_i$ parameters have been sampled from this prior, they can be transformed back into mixing ratios (following equations \ref{eq:sum1}, \ref{eq:xi_i} and \ref{eq:g_x}), and used to calculate the model spectrum:
\begin{equation*}
    X_i = \frac{e^{\xi_i}}{\sum_{j=1}^{n}e^{\xi_j}}.
\end{equation*}

The CLR parameterisation results in identical priors for the mixing ratios of each chemical species in the model, as shown in Figure \ref{fig:priors}. These priors are spiked towards higher abundances given the necessity that at least one species must have a large abundance (as noted by \citealt{Benneke2012}). The resulting parameter space therefore considers all chemical species in the model equally, allowing any of them to be the dominant atmospheric species.

\subsubsection{Atmospheric opacity}
\label{sec:ret_opac}

\begin{figure}
    \centering
    \includegraphics[width=0.49\textwidth]{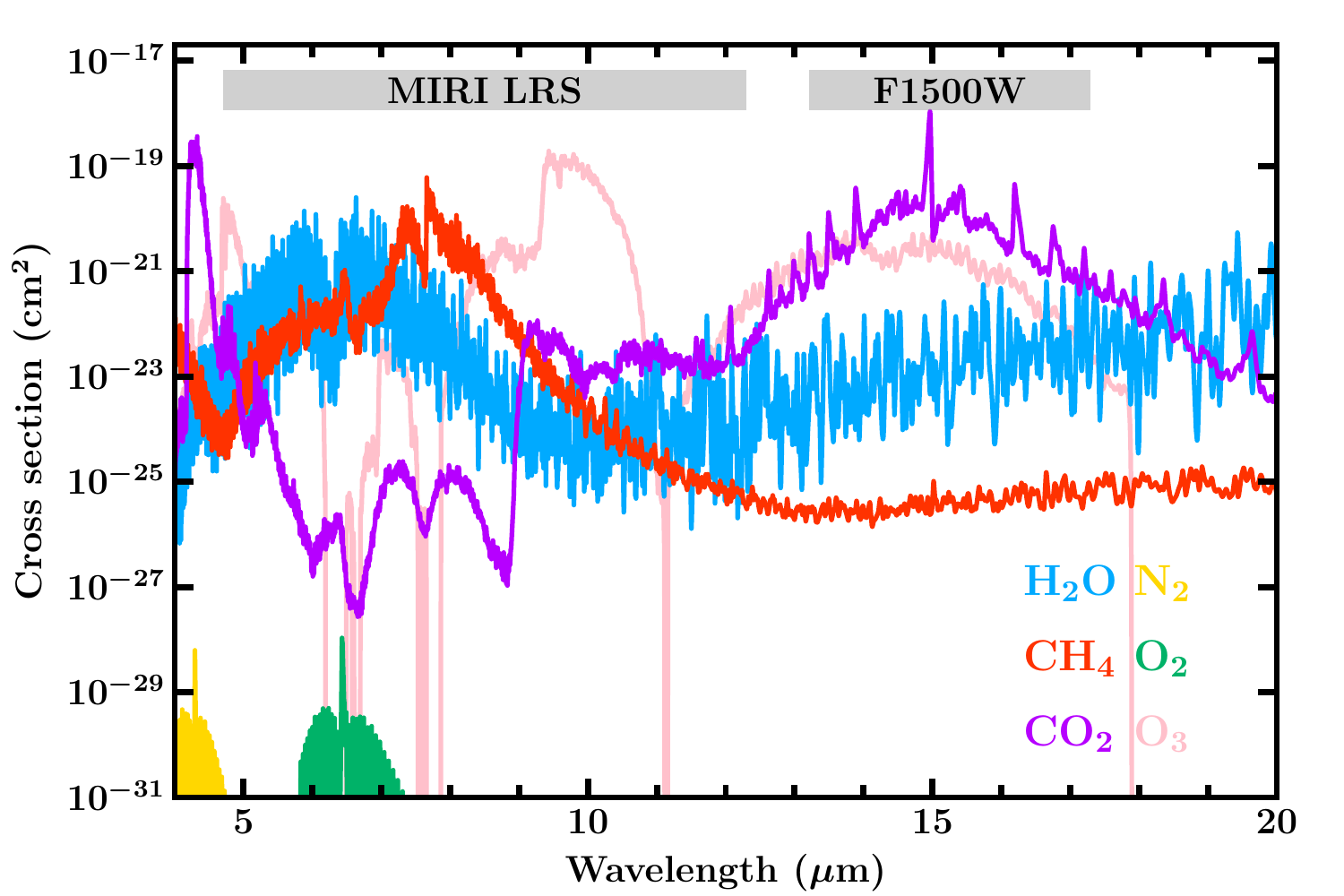}
    \caption{Key sources of opacity in the mid-infrared for secondary atmospheres, at a pressure of 1~bar and temperature of 500~K. The wavelength ranges of MIRI LRS and the MIRI F1500W photometric band are shown by thick grey lines.}
    \label{fig:xsec}
\end{figure}

Since rocky exoplanet atmospheres may be dominated by high mean-molecular-weight (high-$\mu$) species, it is important to include all relevant sources of opacity for such species, including collision-induced absorption (CIA) and Rayleigh scattering. In our models, we therefore include the latest CO$_2$-CO$_2$ and N$_2$-N$_2$ CIA opacities from the HITRAN database \citep{Karman2019}, as well as H$_2$-H$_2$ CIA opacity \citet{Richard2012}. We further include Rayleigh scattering due to CO$_2$, H$_2$O and H$_2$ \citep[e.g.][]{Malik2019}. We note that the CIA data currently available is limited in temperature and wavelength. In this work, we therefore assume no CIA opacity outside the wavelength ranges available, and set the opacity at temperatures outside these limits to those at the boundary temperatures. Given the wide range of temperatures known for rocky exoplanets, future work will be required to more accurately assess the role of CIA opacity on their atmospheric spectra.

We also include molecular opacity due to species expected to be important in rocky exoplanet atmospheres which also have spectral features in the JWST/MIRI spectral range ($\sim$5-30~$\mu$m), i.e. CO$_2$, H$_2$O, CH$_4$, O$_2$, O$_3$, and N$_2$. These species are known to be important in the atmospheres of the solar system terrestrial planets and/or may dominate the atmospheres of rocky exoplanets in the temperature range considered here, i.e. $T_{\rm eq}\sim 400-800~K$ \citep[e.g.][]{Herbort2020,Hu2020,Thompson2021,Wunderlich2021}. While CO is also thought to be important in such atmospheres, it does not have significant features in the MIRI spectral range and we therefore exclude it from our present retrievals. In order to simulate a realistic, agnostic retrieval analysis, we include all of the opacity sources listed here in each of the retrievals in this work, regardless of the `true' input composition.

The molecular cross sections for the species we include are calculated as described in \citet{Gandhi2017} using line lists from the ExoMol, HITEMP and HITRAN databases (CO$_2$ and H$_2$O: \citealt{Rothman2010}, CH$_4$: \citealt{Yurchenko2013,Yurchenko2014a}, O$_2$ and O$_3$: \citet{Rothman2013}, N$_2$: \citealt{Barklem2016,Western2018}). As discussed in \citet{Gandhi2017}, these cross sections are pressure broadened using the parameters for air broadening. Ideally, the pressure broadening would reflect the true composition of the atmosphere for each atmospheric model computed in the retrieval. However, given the wide range of possible atmospheric compositions for rocky exoplanets, this approach would be very cumbersome \citep[e.g.][]{Scheucher2020} and computationally expensive in the context of an atmospheric retrieval, in which a wide range of compositions is explored. Furthermore, the relevant line broadening coefficients are not all currently known \citep[e.g.][]{Fortney2019}. Further work on the pressure broadening of molecular opacities in rocky exoplanet atmospheres will be needed in order to improve these models and retrievals in the future. 

Figure \ref{fig:xsec} shows the molecular opacities included in the retrieval as a function of wavelength, as well as the spectral coverage of JWST/MIRI's Low Resolution Spectroscopy (LRS) mode and F1500W photometric band. Some species have overlapping molecular absorption in the MIRI spectral range and may potentially lead to degeneracies in the retrieved abundances, depending on the data quality and spectral coverage. For example, H$_2$O and CH$_4$ have a somewhat similar cross section profile in the MIRI LRS spectral range. Similarly, CO$_2$ and O$_3$ both have strong features at $\sim$9~$\mu$m. In section \ref{sec:case_studies}, we consider atmospheric compositions ranging from CO$_2$-rich to H$_2$-O-rich, and find that H$_2$O and CO$_2$ can both be detected in the thermal emission spectrum despite these degeneracies. 

\subsubsection{Atmospheric detections}
\label{sec:ret_detect}

The Bayesian evidence of a retrieved model fit can be use to statistically and robustly compare different retrieval models \citep[e.g.][]{Trotta2008}. The Nested Sampling algorithm used in \hydrar calculates the Bayesian evidence of the retrieved model fit and therefore allows such comparisons to be made. Bayesian model comparison is commonly used to assess the statistical significance of molecular detections by comparing retrieval models which include/exclude a certain molecule \citep[e.g.][]{Benneke2013,Gandhi2018}. Here, we consider how such model comparison can be used to assess the presence of an atmosphere or specific molecules by comparing models with/without the presence of molecular features. 

For two different retrieval models, $a$ and $b$, the ratio of their Bayesian evidences (i.e. the Bayes factor), 
\begin{equation*}
    \mathcal{B}_{a,b} = \frac{p(\mathrm{data}|\mathrm{model}\,a)}{p(\mathrm{data}|\mathrm{model}\,b)},
\end{equation*}
provides a measure of how likely model $a$ is relative to model $b$. For example, $\ln(\mathcal{B}_{a,b})=$ 1.0, 2.5 or 5.0 suggests weak, modest and strong evidence for model $a$ over model $b$, respectively \citep{Trotta2008}. The Bayes factor can further be converted to a `sigma' value to represent the confidence of the detection \citet{Benneke2013}. The comparison of Bayes factors can be used to evaluate the significance of specific molecular detections \citep[e.g.][]{Benneke2013,Gandhi2018}. To do this, a retrieval which includes all model parameters is compared to a model which includes all model parameters apart from the abundance of the molecule in question. 

The presence of an atmosphere on a transiting exoplanet can be inferred in several ways \citep[e.g.][]{Koll2019b,Mansfield2019}, including the detection of absorption features in its thermal emission spectrum. In order to calculate the significance of such an atmospheric detection, we compare each of our retrievals in Sections \ref{sec:validation} and \ref{sec:case_studies} to a featureless blackbody retrieval model, whose only parameter is the surface temperature. The resulting Bayes factor therefore indicates the confidence with which atmospheric absorption is detected. We note that surface reflection may lead to a bare planet having spectral features despite the lack of an atmosphere \citep[e.g.][]{Hu2012}. However, such features are typically expected to be small; comparison to a blackbody spectrum thus provides a good first-order estimate of whether an atmosphere is present.

\subsection{Self-Consistent Atmospheric Model \& Simulated Data}
\label{sec:methods_spectra}

In order to test the observability of known rocky exoplanets, we first use self-consistent 1-D atmospheric models to simulate their thermal emission spectra and then use \hydrar to assess the observations needed for robust chemical detections. We use an adaptation of the \textsc{genesis} self-consistent atmospheric model \citep{Gandhi2017,Piette2020a,Piette2020c} to do this. \textsc{genesis} self-consistently solves radiative-convective, hydrostatic and thermochemical equilibrium to find the steady-state $P$-$T$ profile and thermal emission spectrum of the atmosphere. We use direct, second-order methods to solve the equation of radiative transfer; as described in \citet{Piette2020c}, we use the Feautrier method \citep{Feautrier1964} in the iterative solution of radiative-convective equilibrium, and the Discontinuous Finite Element method \citep{Castor1992} combined with Accelerated Lambda Iteration for the final calculation of the spectrum. 

At the base of the atmosphere, we assume a surface pressure of 10~bar and a small internal flux corresponding to an internal temperature of 10~K, i.e. somewhat comparable to the internal temperatures of the terrestrial planets in the solar system \citep[e.g.][]{Zharkov1983,Davies2010,Parro2017}. We note that in the retrievals shown in sections 2.3 and 4, we assume a fixed surface pressure equal to that used here to generate the simulated data (i.e. 10~bar). The retrieval is not sensitive to the choice of surface pressure, as long as it is deep enough to encompass the photosphere. We demonstrate this in Appendix \ref{sec:highP_appendix} for the validation retrieval of Trappist-1~b shown in Section \ref{sec:validation}.

We model the stellar irradiation for GJ~1132~b using a Kurucz stellar model \citep{Kurucz1979,Castelli2003} corresponding to the stellar properties listed in Table \ref{tab:props}. For the stellar spectra of Trappist-1 and LHS~3844, however, we use PHOENIX models \citep{Phoenix2013} as their effective temperatures ($\sim$2500~K and $\sim$3000~K, respectively) are significantly below the coolest temperature considered in the Kurucz models (3500~K). We further assume full day-night energy redistribution, i.e. a flux redistribution factor of $f$=0.25 according to the notation of \citet{Burrows2008a}.

In section \ref{sec:case_studies}, we focus on three key atmospheric compositions ranging from CO$_2$-rich to H$_2$O rich. These are: (i) a Venus-like composition with 97\% CO$_2$, 2.9\% N$_2$ and 0.1\% H$_2$O by volume, (ii) a 50\%~CO$_2$, 50\%~H$_2$O composition (by volume) and (iii) a 100\%~H$_2$O composition. We include molecular opacity from each of these species as well as CO$_2$-CO$_2$ and N$_2$-N$_2$ CIA, and Rayleigh scattering due to CO$_2$ and H$_2$O (see Section \ref{sec:ret_opac}). For simplicity, we assume constant-with-depth chemical abundances. We note that a range of other species (e.g. NH$_3$, HCN, SO$_2$ or H$_2$S) may also be present in rocky exoplanet atmospheres, for example due to photochemistry or outgassing \citep[e.g.][]{Moses2014,Herbort2020,Thompson2021,Wunderlich2021,Yu2021}. However, for simplicity we focus here on CO$_2$/H$_2$O-dominated compositions. For each of the planets modelled in Sections \ref{sec:validation} and \ref{sec:case_studies}, we list the planetary and stellar parameters used in Table \ref{tab:props}.

From the model thermal emission spectra calculated using \textsc{genesis}, we simulate both MIRI LRS spectra and MIRI photometry. For the MIRI LRS spectra, we bin the model spectrum to the pixel resolution and convolve it to a resolution of $R\sim100$ (i.e. close to the instrument resolution) using a Gaussian kernel. The Gaussian kernel has fixed full width at half maximum (FWHM), chosen such that FWHM= $\lambda/100$ at the centre of the MIRI LRS spectral range, i.e. at $\lambda=8.5$~$\mu$m. We calculate the uncertainties on the simulated LRS data using \textsc{PandExo} \citep{Pandexo} using a noise floor of 20~ppm at native resolution, and add this as random Gaussian noise to the simulated data. For the simulated MIRI photometry, we bin the spectra using the instrument response functions for each band \citep{Glasse2015} and assume nominal single-eclipse uncertainties of 100~ppm, i.e. a conservative estimate of the uncertainties expected for the targets considered in section \ref{sec:case_studies} (e.g. \citealt{Lustig-Yaeger2019}). As with the simulated MIRI LRS data, we add this uncertainty as random Gaussian noise to the simulated photometry data. In this work, we use the F1500W photometric band at $\sim$15~$\mu$m as this probes opacity due to CO$_2$ and H$_2$O, as shown in Figure \ref{fig:xsec}.

\begin{table*}
  \centering
  \caption{Planetary and stellar parameters for the rocky exoplanets modelled in this work. Values in parentheses are those used when data is unavailable or rounded values used for a \textsc{Phoenix} stellar model in the case of Trappist-1 and LHS~3844. Stellar K magnitudes are from 2MASS \citep{2MASS2003}.}
    \begin{tabular}{|c|c|c|c|c|c|c|c|c|c|c|}
    \hline
    \textbf{Planet} & \boldmath{}\textbf{$R_\mathrm{p}$ ($R_\oplus$)}\unboldmath{} & \boldmath{}\textbf{$M_\mathrm{p}$ ($M_\oplus$)}\unboldmath{} & \boldmath{}\textbf{$T_\mathrm{eq}$ (K)}\unboldmath{} & \boldmath{}\textbf{$a$ (au)}\unboldmath{} & \boldmath{}\textbf{$R_\mathrm{s}$ ($R_\odot$)}\unboldmath{} & \boldmath{}\textbf{$T_\mathrm{eff}$ (K)}\unboldmath{} & \boldmath{}\textbf{$\log(g_\mathrm{s}/\mathrm{cgs})$}\unboldmath{} & \textbf{[Fe/H]} & \textbf{K mag} & \textbf{Refs} \bigstrut\\
    \hline
    \hline
    Trappist-1 b & 1.086 & 0.85  & 400   & 0.01111 & 0.117 & 2559 (2500) & 5.2 (5.0) & 0.04 (0.0) & 10.296 & 1 \bigstrut\\
    \hline
    GJ 1132 b & 1.130 & 1.66  & 585   & 0.0153 & 0.2105 & 3270  & 4.881 & -0.12 & 8.322 & 2,3,4 \bigstrut\\
    \hline
    LHS 3844 b & 1.303 & 2.25$^*$ & 807   & 0.00622 & 0.189 & 3036 (3000)  & 5.06 (5.0) & (0.0) & 9.145 & 5 \bigstrut\\
    \hline
    \end{tabular}%
    \begin{tablenotes}
    \item $^*$Calculated mass assuming an Earth-like composition. 
    \item \footnotesize{{\bf References:} 1.~\citet{Gillon2017}, 2.~\citet{Bonfils2018}, 3.~\citet{Southworth2017}, 4.~\citet{Berta-Thompson2015}, 5.~\citet{Vanderspek2019}.} 
    \end{tablenotes}
    \label{tab:props}%
\end{table*}%


\subsection{Validation of Retrieval Framework}
\label{sec:validation}
\begin{figure*}
    \centering
    \includegraphics[width=0.45\textwidth]{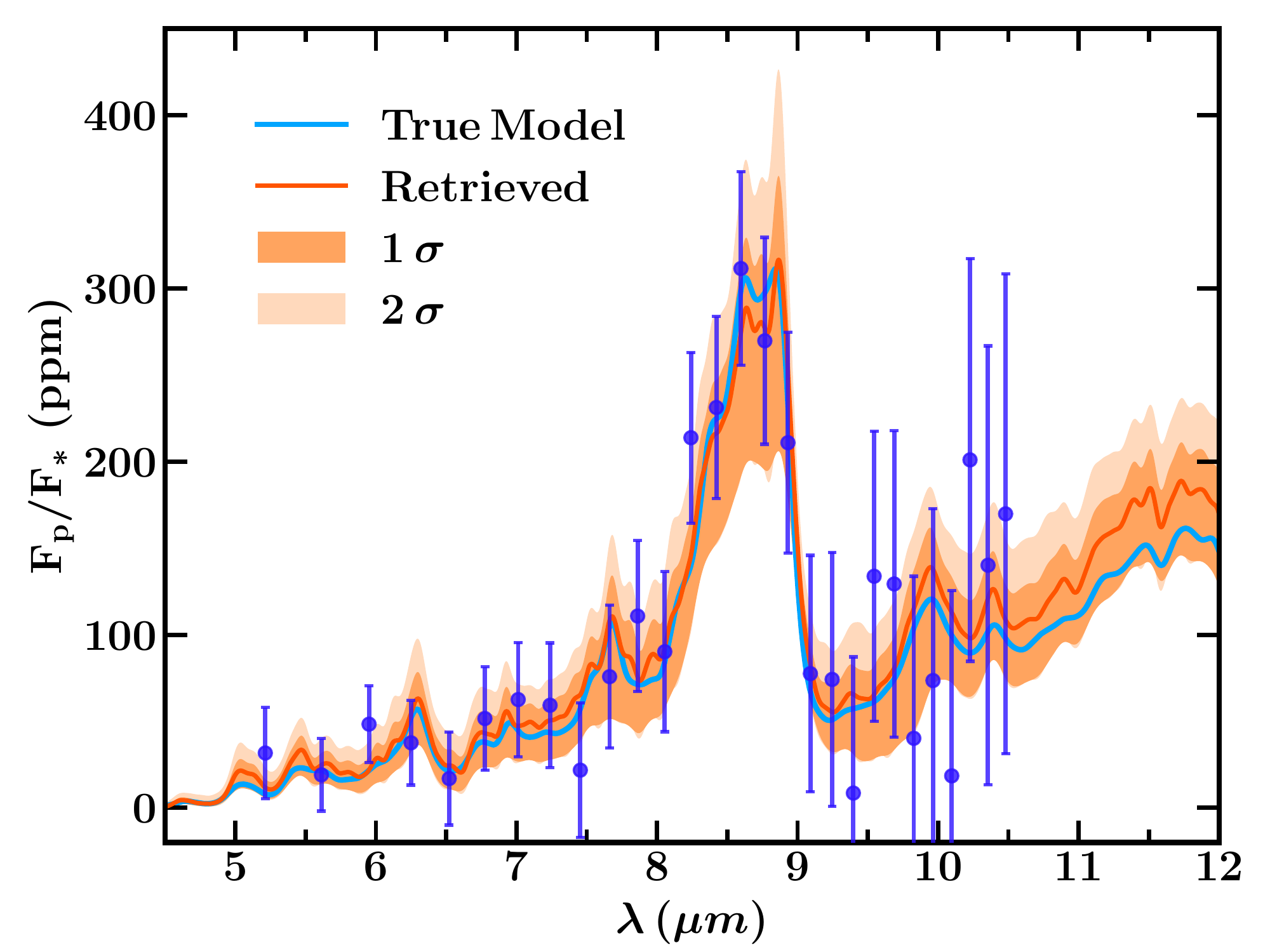}
    \includegraphics[width=0.45\textwidth]{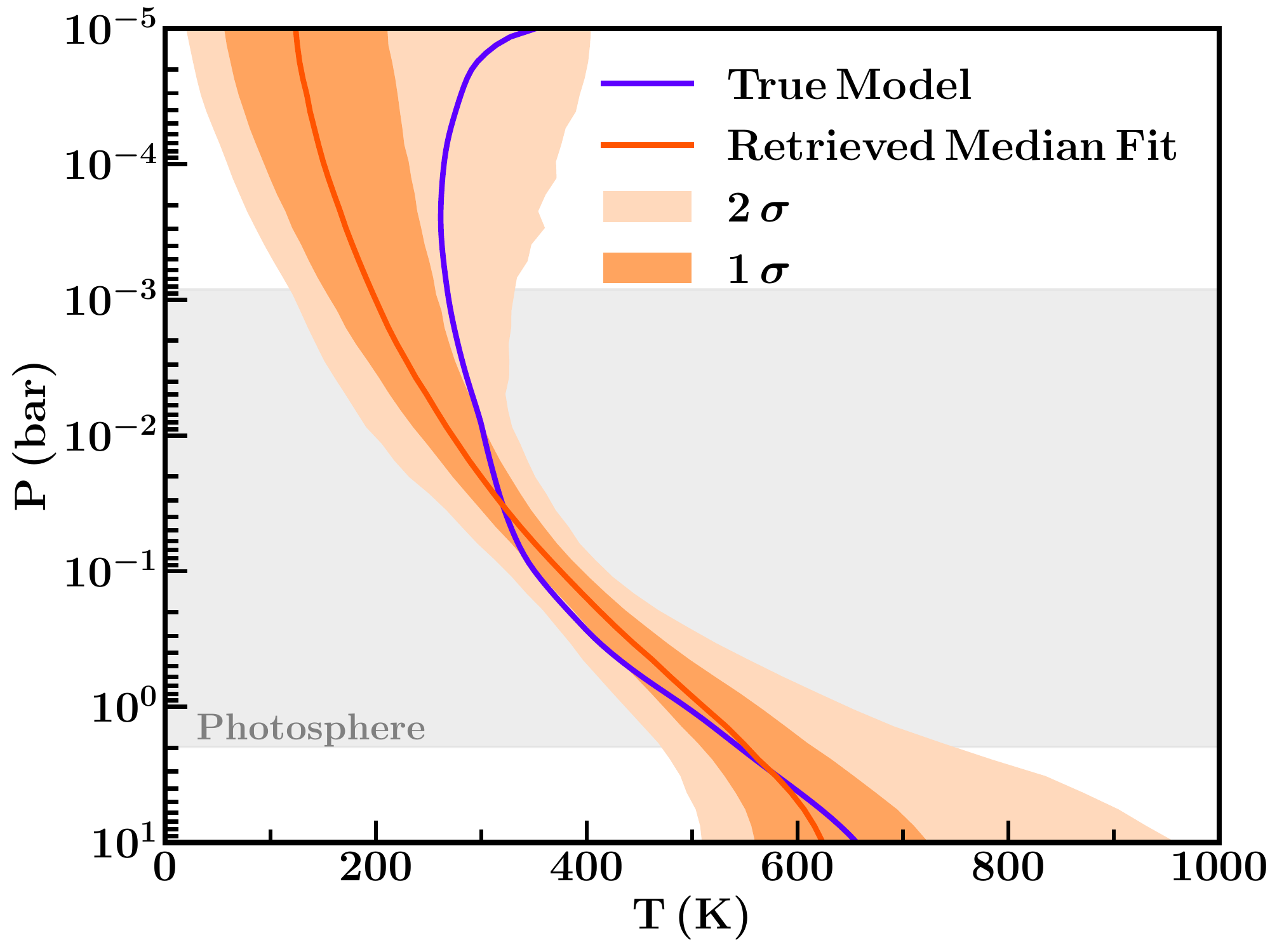}
    \includegraphics[width=0.95\textwidth]{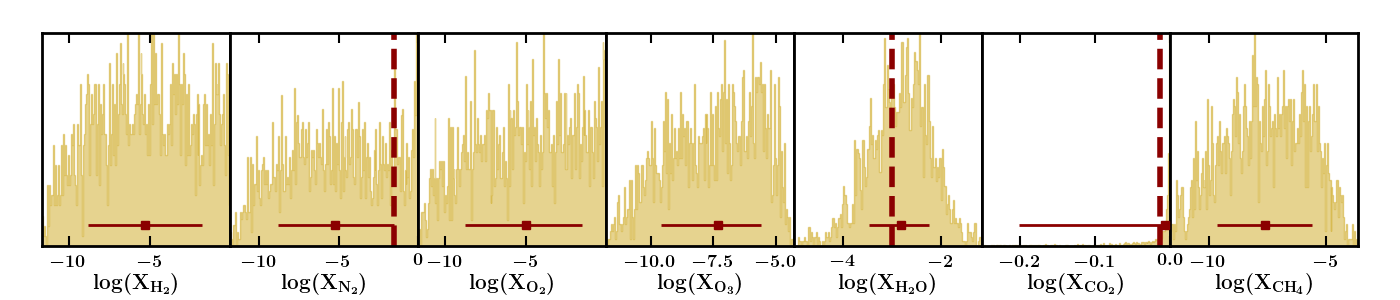}
    \caption{Retrieved emission spectrum, temperature profile and chemical abundances for simulated data of Trappist-1~b, assuming a Venus-like composition. The data are simulated for 30 MIRI LRS eclipses \textit{Top left}: self-consistently modelled `true' spectrum for Trappist-1~b (blue line) and corresponding simulated MIRI LRS data (purple markers and error bars). The median retrieved spectrum and 1$\sigma$/2$\sigma$ contours are shown by the orange line and dark/light orange shading, respectively. \textit{Top right}: self-consistent `true' temperature profile (purple line) and median retrieved temperature profile and 1$\sigma$/2$\sigma$ contours (orange line and dark/light orange shading, respectively). The shaded grey region shows the range of the 5-20$\mu$m photosphere assuming R$\sim$100. \textit{Bottom}: Retrieved posteriors for the chemical abundances. `True' input values are shown by vertical dashed lines. Maroon markers and error bars show the median retrieved abundances with 1$\sigma$ error bars. CO$_2$ and H$_2$O are detected at 5.51$\sigma$ and 4.90$\sigma$ statistical significance, respectively.}
    \label{fig:validation}
\end{figure*}

We validate the retrieval framework by applying it to simulated data for Trappist-1~b. We further compare our results to \citet{Lustig-Yaeger2019}, who have assessed the atmospheric observability of the Trappist-1 planets. We model the spectrum of Trappist-1~b assuming a Venus-like composition and simulate its MIRI LRS spectrum, as described in Section \ref{sec:methods_spectra}. We simulate data uncertainties assuming 30 eclipses. The self-consistent temperature profile and thermal emission spectrum obtained (including simulated data) are shown in Figure \ref{fig:validation}. Note that the simulated MIRI LRS data shown in Figure \ref{fig:validation} have been binned for clarity, but the data used for the retrieval are at the native resolution.

The retrieved spectrum and temperature profile are also shown in Figure \ref{fig:validation}, alongside the posterior distributions for the chemical abundances in the model. 2D marginalised posterior probability distributions are shown in Appendix \ref{sec:corner_plot_appendix}. Both CO$_2$ and H$_2$O (the only spectrally active species in the self-consistent model) are detected with strong statistical confidence, at 5.51$\sigma$ and 4.90$\sigma$, respectively. Correspondingly, the retrieved spectrum is able to fit the CO$_2$ absorption feature at $\sim$9$\mu$m as well as the broad absorption feature due to H$_2$O around $\sim$6~$\mu$m. This result is consistent with \citet{Lustig-Yaeger2019}, who find that $\sim$30 eclipses with MIRI LRS are able to significantly rule out a featureless emission spectrum for Trappist-1~b, in the case of a clear 10~bar CO$_2$ atmosphere. 

We further find that the retrieval is able to accurately fit the photospheric temperature profile with very good precision, within $\lesssim$200~K at 2$\sigma$. The retrieved $P$-$T$ profile in Figure \ref{fig:validation} is able to reproduce the `true' input temperature profile to within 2$\sigma$ throughout the atmosphere, with a tighter fit at photospheric pressures and larger uncertainty outside the photospheric range (as expected, since these regions have little effect on the observable spectrum). Thermal emission observations of rocky exoplanets therefore have the potential to place exquisite constraints on the thermodynamic conditions in their atmospheres.

\section{Target Selection}
\label{sec:targets}

In this Section, we evaluate the observability of known rocky exoplanets in thermal emission, and identify optimal candidates for atmospheric characterisation. Since the planet-star contrast is greater at longer wavelengths, we specifically assess the planets' observability in the mid-infrared with MIRI LRS, i.e. in the spectral range 5-12~$\mu$m. We begin by considering all known exoplanets in the NASA Exoplanet Archive\footnote{exoplanetarchive.ipac.caltech.edu} with masses $<10 M_\oplus$ and radii $<2 R_\oplus$ whose stars have K magnitudes $<13$. We further include LHS~3844~b, which does not have a measured mass but represents an ideal target for rocky exoplanet characterisation.

In order to identify optimal targets for thermal emission observations, we first assess these targets by considering their planetary and stellar spectra to be blackbodies. For the planetary temperature, we calculate the equilibrium temperature of the planet assuming zero albedo and full day-night energy redistribution, i.e.
\begin{equation*}
    T_\mathrm{p} = \sqrt{\frac{R_\mathrm{s}}{2a}}T_\mathrm{eff},
\end{equation*}
where $R_\mathrm{s}$ and $T_\mathrm{eff}$ are the stellar radius and effective temperature, respectively, and $a$ is the semi-major axis. We then calculate the planet-star flux ratio for each system, 
\begin{equation*}
    \frac{F_\mathrm{p}}{F_\mathrm{s}} = \frac{R_\mathrm{p}^2B_\lambda(T_\mathrm{p})}{R_\mathrm{s}^2B_\lambda(T_\mathrm{eff})},
\end{equation*}
where $R_\mathrm{p}$ is the planetary radius and $B_\lambda$ is the Planck function. We filter the targets according to $F_\mathrm{p}/F_\mathrm{s}$ in three steps:
\begin{enumerate}
    \item Filter out any targets for which $F_\mathrm{p}/F_\mathrm{s}<$10~ppm at 12~$\mu$m
    \item Recalculate $F_\mathrm{p}/F_\mathrm{s}$ using a Phoenix spectrum for the star (still assuming a blackbody spectrum for the planet), and use \textsc{PandExo} \citep{Pandexo} to calculate the uncertainty in the MIRI LRS spectrum. 
    \item Calculate the number of eclipses needed, $E_{\rm S/N=3}$, to achieve a signal-to-noise (S/N) of 3 anywhere in the MIRI~LRS range assuming a resolution of $R\sim$10. We use $E_{\rm S/N=3}$ as a metric for the observability of the target.
\end{enumerate}

By evaluating $E_{\rm S/N=3}$ at a resolution of $R\sim$10, this metric quantifies the spectroscopic observability of thermal emission from rocky exoplanets and indicates which planets may be amenable to atmospheric characterisation. Since CO$_2$ and H$_2$O have broad prominent features in the MIRI~LRS spectral range, $R\sim$10 is a suitable resolution to identify the presence of such species in the atmospheres of rocky exoplanets. For example, we find that the promising rocky exoplanet target GJ~1132~b has $E_{\rm S/N=3}$=1.32, and in section \ref{sec:case_studies} we show that 8 secondary eclipses with MIRI~LRS are sufficient to constrain a CO$_2$-rich atmosphere. Conversely, fewer eclipses would be required to make photometric detections of the planetary thermal emission, e.g. as discussed by \citet{Koll2019b}.

\begin{figure*}
    \centering
    \includegraphics[width=\textwidth]{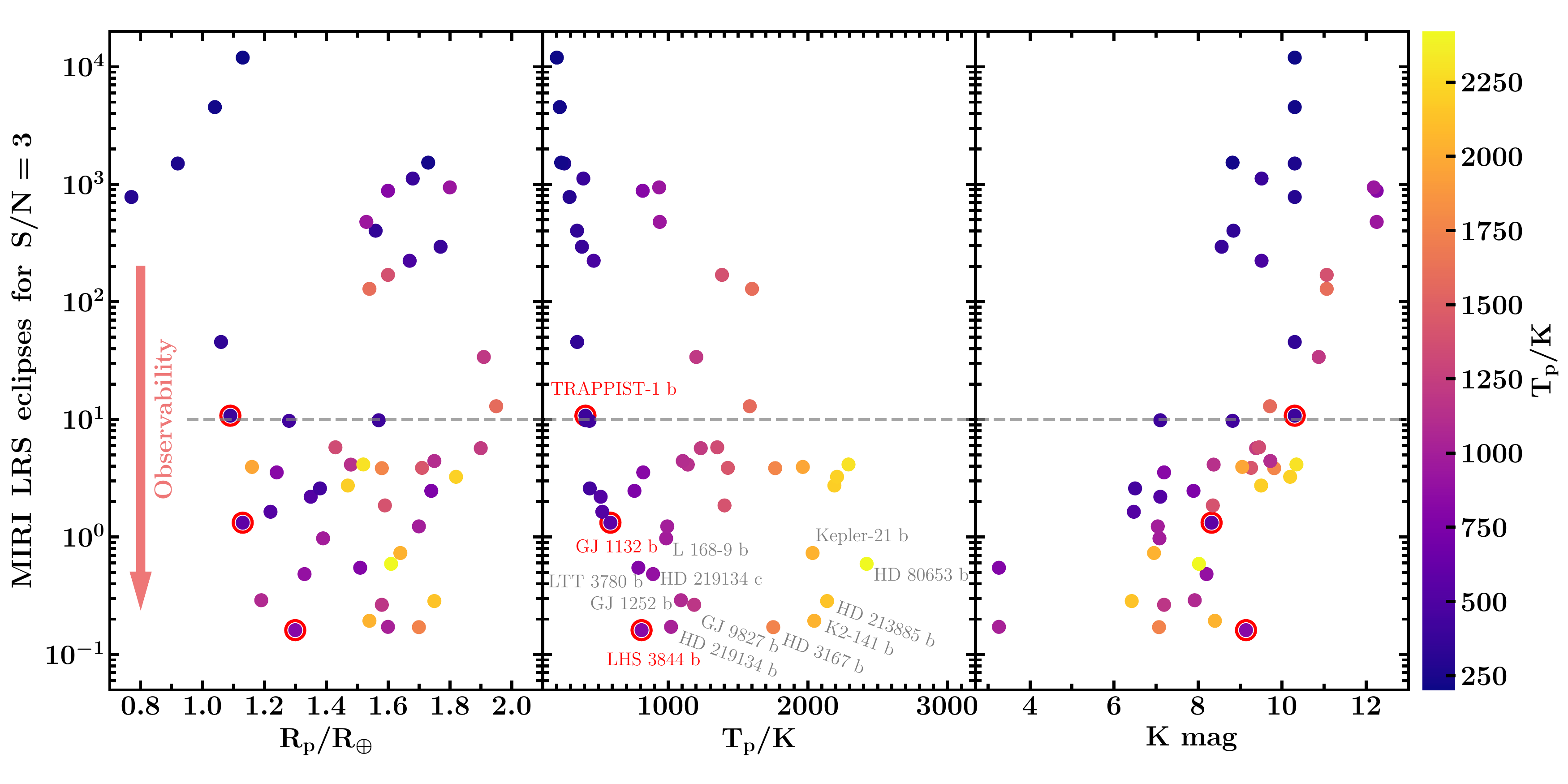}
    \caption{Observability of known rocky planets in thermal emission as a function of planet radius (R$_{\rm p}$, left panel), equilibrium temperature (T$_{\rm p}$, centre panel) and K magnitude of the host star (right panel). Observability is defined as the number of secondary eclipses with MIRI LRS required to detect the thermal emission of the planet, $F_\mathrm{p}/F_\mathrm{s}$, at S/N=3 assuming a resolution of R$\sim$10. Note that non-integer numbers of eclipses would need to be rounded up as partial transits would not contribute as much to the S/N. The targets shown here are chosen such that $F_\mathrm{p}/F_\mathrm{s}>$10~ppm in the MIRI LRS spectral range (i.e 5-12~$\mu$m). In each panel, the markers for each planet are coloured according to planet equilibrium temperature (see colourbar). The case studies LHS~3844~b, GJ~1132~b and Trappist-1~b are circled in red in each panel, for reference. The horizontal dashed line denotes 10 eclipses required with MIRI LRS for a S/N=3. The planets below this line represent optimal targets for atmospheric characterisation in thermal emission with JWST MIRI, and their properties are listed in Table \ref{tab:best_targets}. Note that 55~Cnc~e is not shown here as it would saturate part of the MIRI LRS detector.}
    \label{fig:targets}
\end{figure*}

Figure \ref{fig:targets} shows the observability of the resulting population of rocky exoplanets as a function of their stellar and planetary parameters. As expected, hotter and larger planets are typically more observable than cooler, smaller ones. Furthermore, the number of eclipses required to achieve S/N=3 with MIRI LRS increases with host star K magnitude. This suggests that the two main obstacles for observing rocky planet atmospheres in thermal emission are the level of thermal emission from the planet itself (which affects the measured signal) and the brightness of the host star (which impacts the observational uncertainty). With MIRI LRS, the thermal emission from several hot rocky planets will be observable in fewer than 10 eclipses at R$\sim$10. However, the characterisation of temperate rocky planet atmospheres will require either more observing time, greater sensitivity than that expected of MIRI, or brighter targets. 

We find $>30$ rocky exoplanets whose thermal emission will be detectable with S/N=3 at $R\sim$10 in fewer than 10 secondary eclipses (Table \ref{tab:best_targets}). We note that, while 55~Cnc~e represents an excellent target for thermal emission observations, the brightness of its host star results in partial saturation of the MIRI~LRS detector and we therefore exclude it from the present list. Of the rocky exoplanets we consider, the most observable is LHS~3844~b; with an equilibrium temperature of $\sim$800~K, it is the only planet in this sample cooler than 1000~K with $E_{\rm S/N=3}<1$. In the intermediate-temperature regime, GJ~1132~b represents an excellent target and is the only exoplanet cooler than 600~K with $E_{\rm S/N=3}<4$. In the temperate regime, Trappist-1~b and L~98-59~d both represent ideal targets, having $E_{\rm S/N=3}\sim$10 and equilibrium temperatures of $\sim$400~K.

Our results are consistent with the findings of \citet{Koll2019b}, who use the emission spectroscopy metric (ESM) of \citet{Kempton2018} to characterise the observability of rocky exoplanets. Concurrent with the trend found in figure \ref{fig:targets}, they find that the observability of rocky exoplanets in thermal emission typically increases with equilibrium temperature. They also predict that TESS will find $\sim$19 exoplanets with radius$<1.5R_\oplus$ and an ESM greater than that of GJ~1132~b (which is considered to be an optimal rocky exoplanet candidate for atmospheric characterisation, e.g. \citealt{Morley2017b}). By considering known exoplanets with masses $<10 M_\oplus$ and radii $<2 R_\oplus$, we find in this work that at least 14 known exoplanets are more observable that GJ~1132~b according to our observability metric. These include the 3 known planets identified by \citet{Koll2019b} as having an ESM greater than that of GJ~1132~b, i.e. HD~213885~b, LHS~3844~b and 55~Cnc~e.

In Section \ref{sec:case_studies}, we perform a more detailed analysis of the three promising targets Trappist-1~b, GJ~1132~b, and LHS~3844~b. These planets each represent ideal targets for their respective equilibrium temperatures, which span the temperate to hot regimes ($\sim$400-800~K). We note that the estimated number of eclipses calculated here for a S/N of 3 is not exact due to the assumption of blackbody spectra, and characterises the ease with which thermal emission from the planet can be observed rather than the number of eclipses required to make chemical detections. However, this analysis does provide a simple metric to assess ideal targets for characterising rocky exoplanet atmospheres, and can guide more detailed evaluations such as those in Section \ref{sec:case_studies}. Indeed, this simple analysis shows that a large number of known rocky exoplanets are suitable targets for atmospheric characterisation in thermal emission with JWST.

\section{Results}
\label{sec:case_studies}

\begin{figure*}
    \centering
    \includegraphics[width=\textwidth]{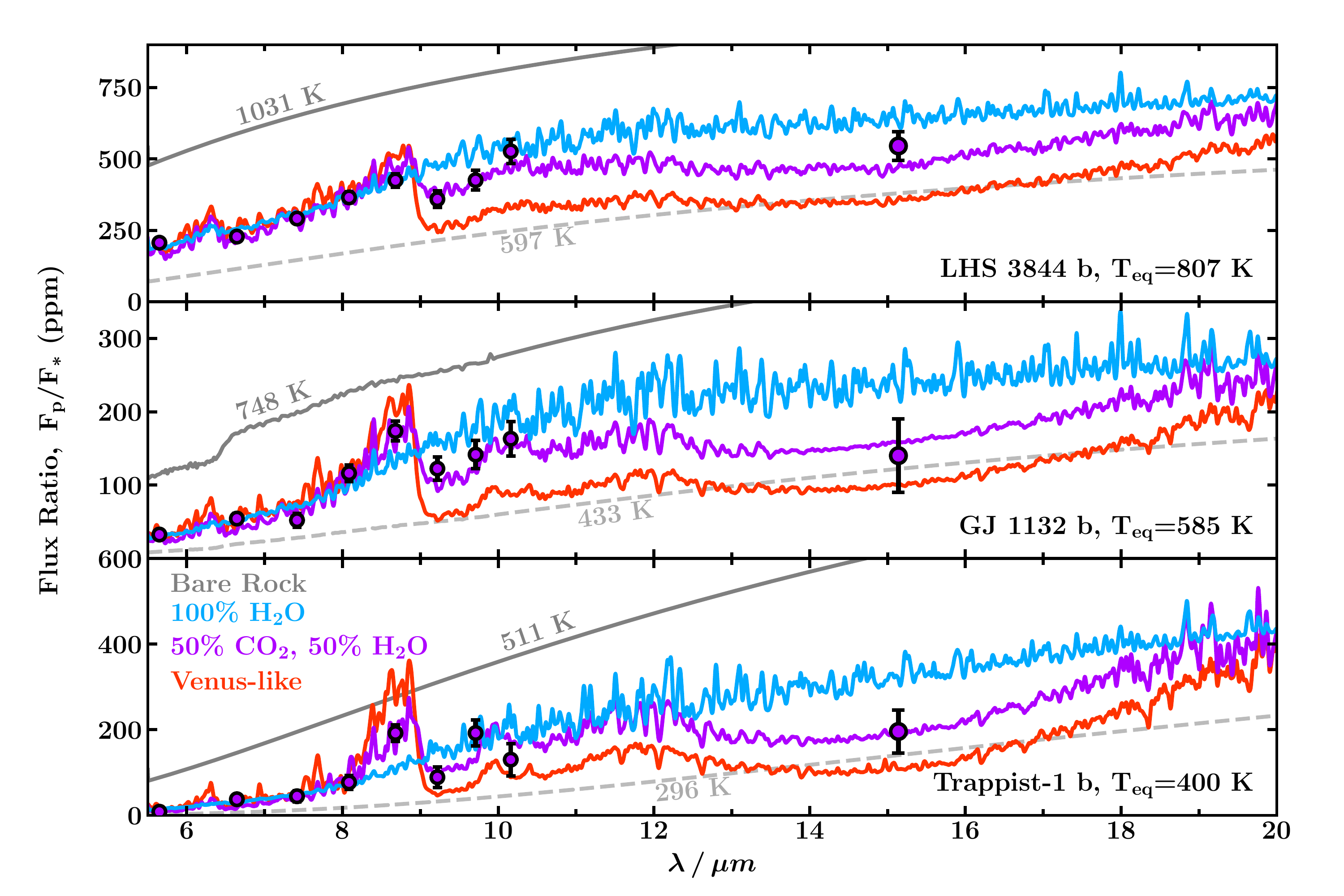}
    \caption{Thermal emission spectra for LHS~3844~b, GJ~1132~b and Trappist-1~b assuming Venus-like (red lines), 50\% CO$_2$/50\% H$_2$O (purple lines) or 100\% H$_2$O (blue lines) compositions. Blackbody spectra corresponding to the temperature expected for a bare rock with no energy redistribution are also shown by solid grey lines. Dashed grey lines show blackbody spectra corresponding to the equilibrium temperature expected for a Venus-like Bond albedo of 0.7. Data points and error bars show simulated MIRI LRS data and photometric data in the F1500W bandpass. The error bars assume 20 MIRI LRS eclipses for LHS~3844~b and GJ~1132~b, and 80 eclipses for Trappist-1~b. Note that the LRS data are binned for clarity. The photometric points have error bars of 50~ppm for all three targets, corresponding to $\sim$4 eclipses each.}
    \label{fig:all_models}
\end{figure*}

\begin{figure*}
    \centering
    \includegraphics[width=0.97\textwidth]{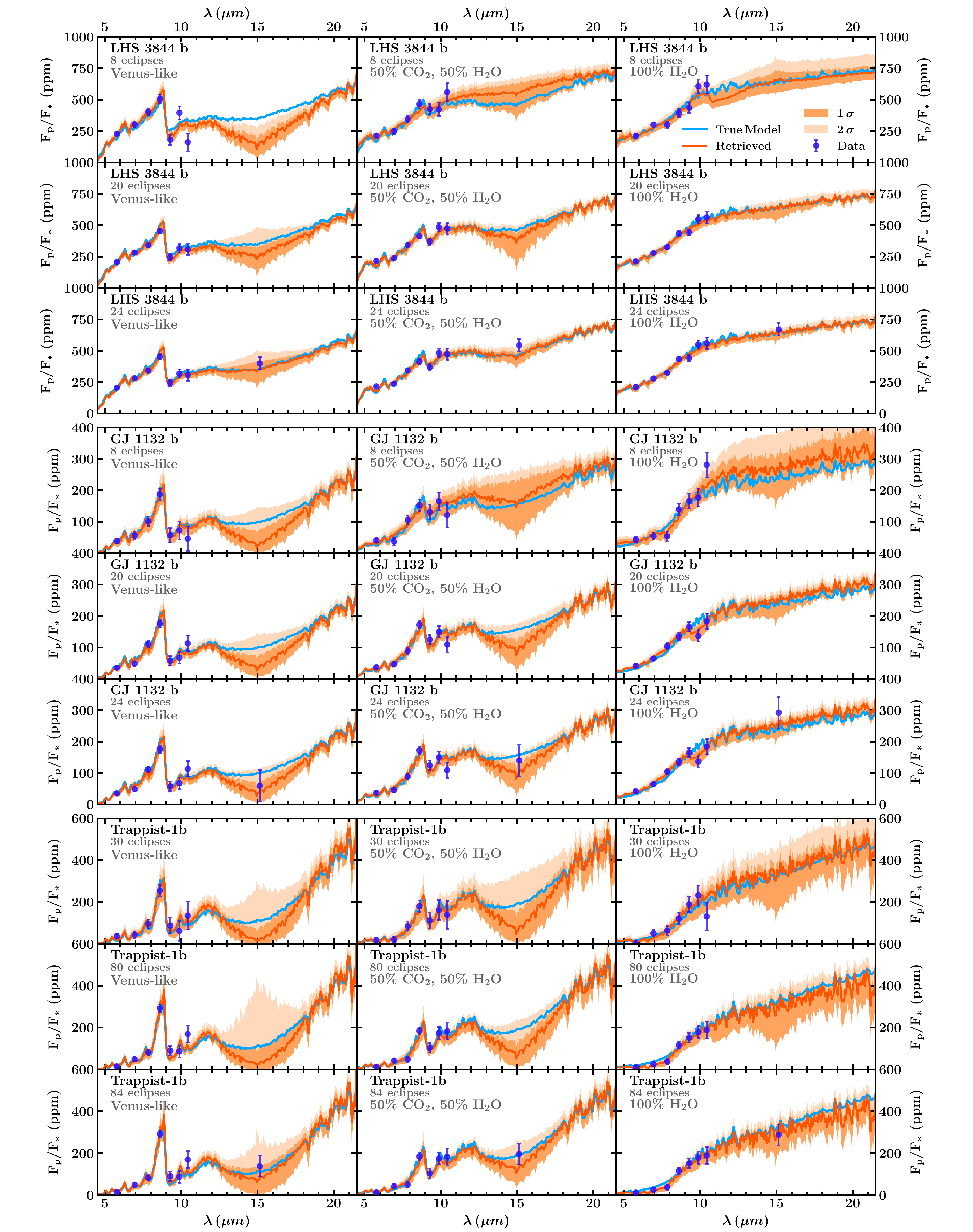}
    \caption{Retrieved thermal emission spectra for LHS~3844~b (top three rows), GJ~1132~b (middle three rows) and Trappist-1~b (bottom three rows) for the simulated data described in section \ref{sec:case_studies} assuming a Venus-like, 50\% CO$_2$/50\% H$_2$O or 100\% H$_2$O composition (left, centre and right columns, respectively). Dark orange lines show the median retrieved spectrum, while dark and light orange shaded regions show 1$\sigma$ and 2$\sigma$ confidence intervals. Purple points and error bars show simulated MIRI LRS and photometric data, as described in Section \ref{sec:case_studies}. The light blue line shows the `true' input spectrum.}
    \label{fig:ret_spectra}
\end{figure*}

We investigate the observability of key molecular species in rocky exoplanet atmospheres using three promising candidates as case studies: Trappist-1~b, GJ~1132~b and LHS~3844~b, which span the $\sim$400-800~K equilibrium temperature range \citep{Berta-Thompson2015,Gillon2016,Vanderspek2019}. In particular, we consider three different atmospheric compositions including a cloud-free Venus-like (CO$_2$-rich) atmosphere, a 50\% CO$_2$/50\% H$_2$O atmosphere and a 100\% H$_2$O atmosphere (see Section \ref{sec:methods_spectra}). For each planet and atmospheric composition, we first model the thermal emission spectrum and simulate JWST MIRI data, as described in Section \ref{sec:methods_spectra}. The model spectra for each case study are shown in Figures \ref{fig:all_models} and \ref{fig:ret_spectra}, including simulated MIRI data. We then perform atmospheric retrievals on this simulated data and assess the detection significances of the key molecular species in each case, as detailed in Section \ref{sec:methods_ret}. We assume full day-night energy redistribution for all our models as this provides a conservative estimate of the observations required to make chemical detections; less efficient heat redistribution would result in a hotter day-side and a deeper secondary eclipse. However, we note that LHS~3844~b's observed phase curve has a strong day-night contrast \citep{Kreidberg2019} which indicates inefficient day-night heat transport and a hotter day-side than assumed here. The observability estimates we provide here are therefore conservative.

\begin{figure*}
    \centering
    \includegraphics[width=0.84\textwidth]{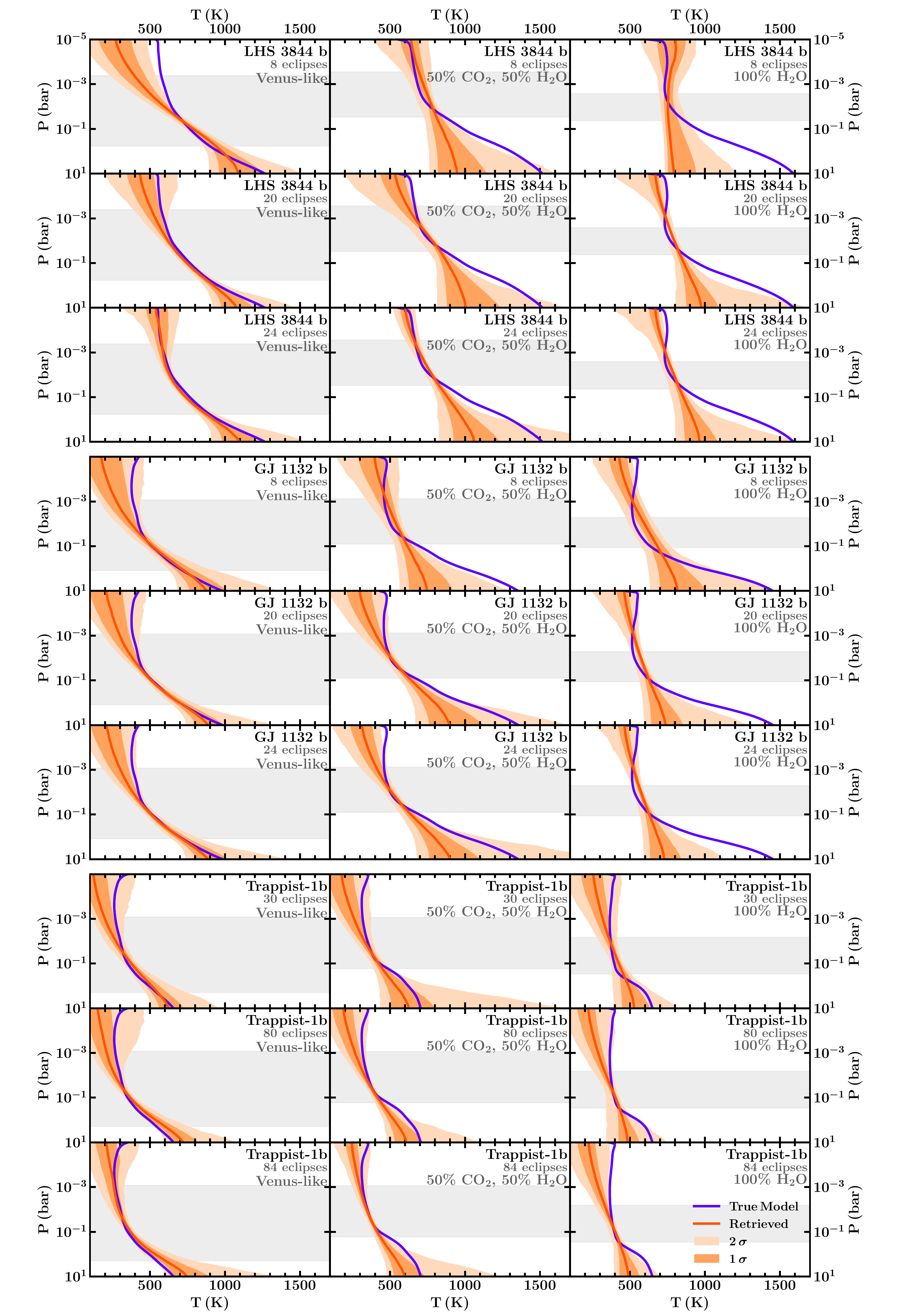}
    \caption{Retrieved temperature profiles for LHS~3844~b (top three rows), GJ~1132~b (middle three rows) and Trappist-1~b (bottom three rows) for the simulated data described in section \ref{sec:case_studies} assuming a Venus-like, 50\% CO$_2$/50\% H$_2$O or 100\% H$_2$O composition (left, centre and right columns, respectively). Dark orange lines show the median retrieved temperature profile, while dark and light orange shaded regions show 1$\sigma$ and 2$\sigma$ confidence intervals. Purple lines show the `true' input temperature profile. Shaded grey regions show the range of the 5-20$\mu$m photosphere assuming R$\sim$100.}
    \label{fig:ret_PT}
\end{figure*}

\begin{figure*}
    \centering
    \vspace{-0.5cm}
    \includegraphics[width=0.8\textwidth]{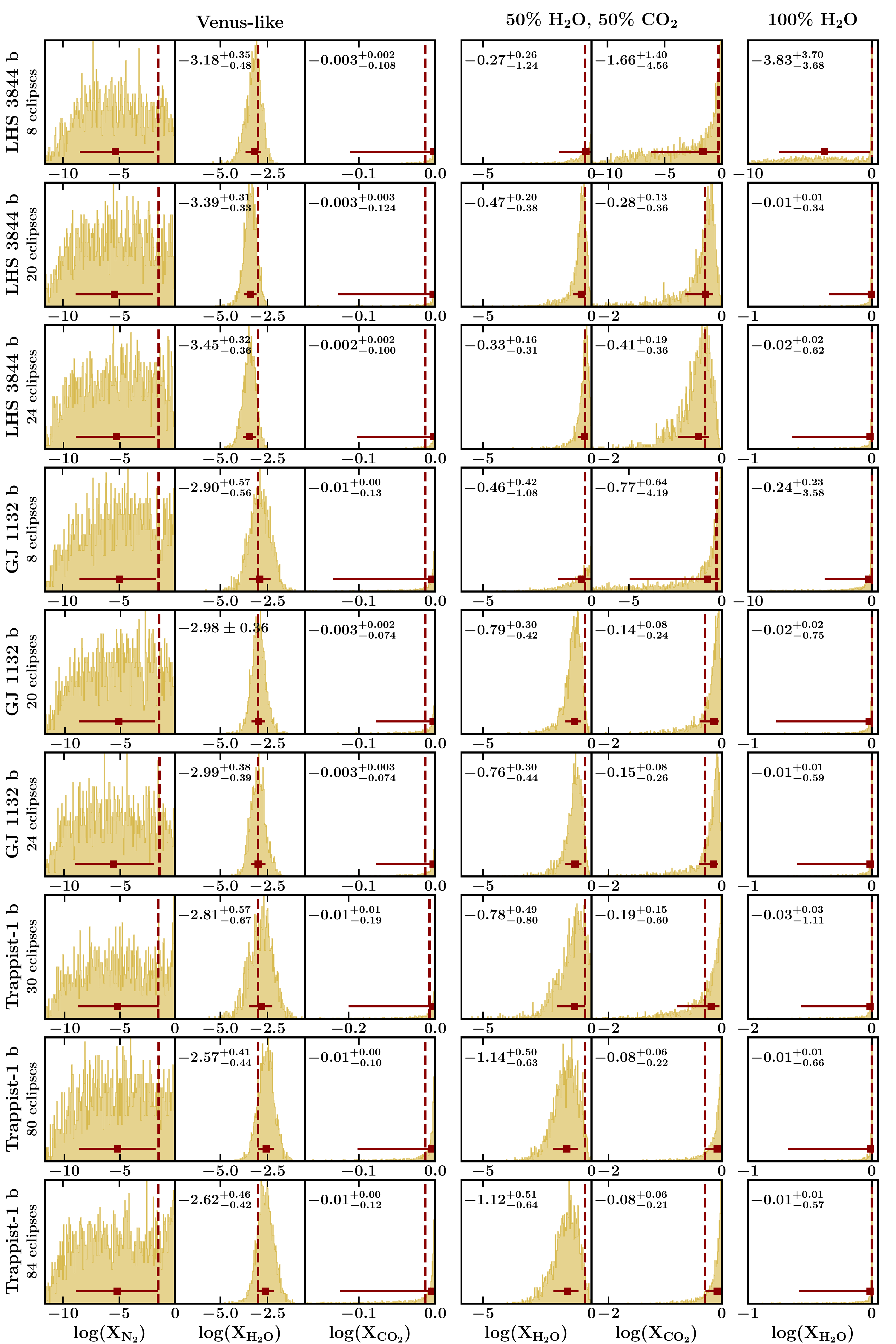}
    \caption{Posterior probability distributions for the species present in each simulated data model, i.e. N$_2$, H$_2$O and CO$_2$ for the Venus-like composition (three left columns), H$_2$O and CO$_2$ for the 50\% CO$_2$/50\% H$_2$O composition (4$^{\rm th}$ and 5$^{\rm th}$ columns) and H$_2$O for the 100\% H$_2$O composition (right column). Note that the retrieval model includes all the species discussed in Section \ref{sec:ret_opac}, but species not present in the simulated data (i.e. those not shown here) are not constrained. Posteriors are shown for the case studies LHS~3844~b (top three rows), GJ~1132~b (middle three rows) and Trappist-1~b (bottom three rows) for the simulated data described in section \ref{sec:case_studies}. `True' input values are shown by vertical dashed lines. Median retrieved abundances and 1$\sigma$ uncertainties are given for each CO$_2$ and H$_2$O posterior distribution, and are shown by maroon markers and error bars for all posteriors. Note that the abundance of N$_2$ is unconstrained in the Venus-like case due to its lack of spectral features in the wavelength range considered. The H$_2$O posteriors for the 100\% H$_2$O composition are typically strongly peaked at 100\%.}
    \label{fig:ret_post}
\end{figure*}

For each of the atmospheres investigated here, we take a hierarchical approach by considering four key science cases to be addressed, in order of increasing scientific complexity: 
\begin{enumerate}
    \item {\bf Comparison to a bare rock with no energy redistribution:} A solid, bare rocky surface is expected to be significantly hotter than an atmosphere with energy redistribution, and may be observationally distinguished from an atmosphere with efficient energy redistribution \citep{Koll2019b}. We consider whether the simulated data are enough to rule out this hot bare rock scenario, indicating energy redistribution e.g. due to an atmosphere or magma ocean.
    \item {\bf Presence of spectral features:} Can the observations statistically rule out a blackbody spectrum? If so, the presence of absorption or emission features confirms the presence of an atmosphere and an atmospheric temperature gradient. 
    \item {\bf Detection of CO$_2$:} CO$_2$ has a sharp spectral feature at $\sim$9~$\mu$m (Figure \ref{fig:xsec}) which can be observed in MIRI LRS spectra. CO$_2$ is therefore one of the easiest molecules to detect in this spectral range if it is sufficiently abundant in the atmosphere.
    \item {\bf Detection of H$_2$O: } With sufficient spectral precision, H$_2$O can be detected and its atmospheric abundance can be measured.
\end{enumerate}
In what follows, we investigate these four science cases for each planet and atmospheric composition. The retrieved thermal emission spectra, temperature profiles and posterior probability distributions for the chemical abundances are shown for each case investigated in Figures \ref{fig:ret_spectra}, \ref{fig:ret_PT} and \ref{fig:ret_post}, respectively.

\subsection{Case Study: LHS 3844 b}

LHS~3844~b is a highly-irradiated rocky planet orbiting a nearby M-dwarf, with a dayside temperature of $\sim$1000~K \citep{Kreidberg2019,Vanderspek2019}. Recent phase curve observations of the planet have revealed a strong day-night temperature contrast and no hotspot shift, ruling out the possibility of an atmosphere with pressure $\gtrsim$10~bar \citep{Kreidberg2019}. Recent observations of LHS~3844~b with transmission spectroscopy are further inconsistent with a clear, H$_2$-rich atmosphere of pressure $\gtrsim$0.1~bar and suggest either a high-$\mu$ atmosphere, a H$_2$-rich atmosphere with high-altitude clouds (cloud-top pressures $<$0.1~bar) or no atmosphere at all \citep{Diamond-Lowe2020}. 

Thermal emission spectroscopy will be required to assess the presence of an atmosphere on LHS~3844~b. Even in the case of a high-$\mu$ atmosphere, which can be prohibitive for transmission spectroscopy, the thermal emission spectrum can be sensitive to molecular absorption/emission features. Here, we assess whether observations with JWST/MIRI could constrain the presence of an atmosphere on LHS~3844~b assuming a cloud-free Venus-like, 50\%CO$_2$/50\%H$_2$O and 100\% H$_2$O composition. For this case study we consider three observing strategies: (i) a shorter observing time strategy with 8 MIRI LRS eclipses, (ii) a longer strategy with 20 MIRI LRS eclipses and (iii) a strategy with 20 MIRI LRS eclipses plus 4 eclipses with the MIRI imager F1500W photometric band. The simulated spectra and data are shown in the top panel of Figure \ref{fig:all_models} and the top three rows of Figure \ref{fig:ret_spectra}. 

From Figure \ref{fig:all_models}, it is clear that the uncertainties on the data using 20 MIRI LRS eclipses and/or the photometry allow the Venus-like, 50\%CO$_2$/50\%H$_2$O and 100\% H$_2$O compositions to be distinguished, especially at wavelengths $\gtrsim$9~$\mu$m. For the longer observing strategies, the data also significantly deviate from a blackbody spectrum for all three compositions considered; for example, the sharp CO$_2$ feature at $\sim$9~$\mu$m can easily be distinguished for the Venus-like and 50\%CO$_2$/50\%H$_2$O models. For the shortest observing strategy of 8 MIRI LRS eclipses, the Venus-like and 100\% H$_2$O cases can also be significantly distinguished from a blackbody spectrum. A bare-rock surface (assuming no energy redistribution, grey spectrum in the top panel of Figure \ref{fig:all_models}) is also evidently rejected by the simulated data for all compositions and observing strategies.

For each atmospheric composition considered, we perform an atmospheric retrieval on the simulated MIRI data described above. The retrieved spectra, temperature profiles and posterior probability distributions for the chemical abundances are shown in the top three rows of Figures \ref{fig:ret_spectra}, \ref{fig:ret_PT} and \ref{fig:ret_post}, respectively, for each case. The `true' input spectra are retrieved within 2$\sigma$ for all three atmospheric compositions, within the spectral range of the data. In all cases, the abundances of H$_2$O and CO$_2$ are accurately retrieved within 2$\sigma$, though in the case of only 8 MIRI LRS eclipses for the 50\%H$_2$O/50\%CO$_2$ and 100\%H$_2$O compositions, these are not detected with statistical significance (Table \ref{tab:sig}). Furthermore, the temperature profile of each atmosphere is accurately retrieved to within $\sim$2$\sigma$ in the photosphere. Outside the photosphere, the observed spectrum does not contain information about the temperature profile and it is therefore expected that the retrieved temperature profile may deviate in this range.

We further evaluate the confidence with which the data reject a blackbody spectrum and the confidence with which CO$_2$ and H$_2$O are detected, as described above and in Section \ref{sec:ret_detect}. With only 8 MIRI LRS eclipses, we find that CO$_2$ and H$_2$O can confidently be detected in the Venus-like case, with detection significances of 5.55$\sigma$ and 4.07$\sigma$, respectively (Table \ref{tab:sig}). Furthermore, while the 50\%~H$_2$O/50\%~CO$_2$ and 100\%~H$_2$O cases are more challenging to characterise with only 8 eclipses, a blackbody spectrum can be rejected at $\lesssim$3$\sigma$ for the 100\%~H$_2$O composition. 8 MIRI LRS eclipses are therefore sufficient to characterise a cloud-free Venus-like composition on LHS~3844~b, or to tentatively detect atmospheric absorption (i.e. reject a blackbody spectrum) in the case of a 100\%~H$_2$O atmosphere.

For the two longer observing time strategies, i.e. with at least 20 MIRI LRS eclipses, stronger constraints can be obtained for all three compositions considered (Table \ref{tab:sig}). For both the Venus-like and 50\%~H$_2$O/50\%~CO$_2$ cases, we find that CO$_2$ and H$_2$O can be detected with $\gtrsim4\sigma$ significance. The 100\%~H$_2$O composition remains more challenging to characterise, but these longer observing time strategies do allow a blackbody spectrum to be rejected at a higher confidence of $>4\sigma$, compared to the 8 eclipse strategy. 20 MIRI LRS eclipses are therefore sufficient to characterise cloud-free, CO$_2$-rich compositions for LHS~3844~b, or to confidently detect atmospheric absorption in the case of a 100\%~H$_2$O atmosphere. We note that the addition of the four photometric eclipses does not significantly improve the retrieved constraints beyond what is achieved with 20 MIRI LRS eclipses alone. Furthermore, for all three observing strategies considered here, the abundances of CO$_2$ and H$_2$O (when detected) can be measured to within $\lesssim0.5$~dex across the compositions we have considered here, i.e. comparable to the precision achieved for chemical abundances in hot Jupiters to date \citep[e.g][]{Sheppard2017,Welbanks2019b}.

\subsection{Case Study: GJ 1132 b}

GJ~1132~b is a small rocky exoplanet with an equilibrium temperature of $\sim$600~K \citep{Berta-Thompson2015}. Thanks to its nearby host star and large planet-star size contrast, this planet is an ideal candidate for thermal emission spectroscopy of warm rocky planets \citep{Morley2017b}. To date, atmospheric observations of GJ~1132~b using transmission spectroscopy have revealed a relatively flat spectrum, consistent with a high-$\mu$ atmosphere \citep{Southworth2017,Diamond-Lowe2018}. Thermal emission spectroscopy will be needed to confirm the presence of this atmosphere and to establish its chemical composition. Here, we consider the JWST/MIRI observations needed to constrain the presence and chemical composition of an atmosphere on GJ~1132~b assuming a cloud-free Venus-like (CO$_2$-rich), 50\%~CO$_2$/50\%~H$_2$O or 100\%~H$_2$O composition.

In order to characterise the observability of GJ~1132~b, we investigate the same three observing strategies used above for LHS~3844~b, i.e.: (i) 8 eclipses with MIRI LRS, (ii) 20 eclipses with MIRI LRS, (iii) 20 eclipses with MIRI LRS plus 4 eclipses with the F1500W photometric band. The simulated spectra and data for these cases are shown in Figure \ref{fig:all_models} and the middle three rows of Figure \ref{fig:ret_spectra}, respectively. For all three observing strategies, the data are clearly distinguishable from the solid bare-rock scenario with no energy redistribution (grey spectrum in Figure \ref{fig:all_models}). For the Venus-like composition, the observational uncertainties for each strategy are further sufficient to distinguish the strong CO$_2$ feature at $\sim$9~$\mu$m. The retrieved spectra, temperature profiles and posterior distributions for each composition and observing strategy are shown in the middle three rows of Figures \ref{fig:ret_spectra}, \ref{fig:ret_PT} and \ref{fig:ret_post}, respectively.

With the assumption of 8 MIRI LRS eclipses, we find that the Venus-like composition is confidently constrained, with 4.7$\sigma$ and 4.4$\sigma$ detections of CO$_2$ and H$_2$O, respectively (Table \ref{tab:sig}). CO$_2$ and H$_2$O are not confidently detected for the 50\%~CO$_2$/50\%~H$_2$O and 100\%~H$_2$O compositions using this strategy, but in the 100\%~H$_2$O case a blackbody spectrum is nevertheless rejected at $>4\sigma$. The spectra are retrieved to within 2$\sigma$ in the spectral rage of the data, while the temperature profiles are also retrieved to within 2$\sigma$ in the photosphere. Furthermore, the abundances of CO$_2$ and H$_2$O are accurately retrieved to within 2$\sigma$. For the Venus-like case, the abundances of CO$_2$ and H$_2$O are constrained within $<0.6$~dex. This relatively short observing time strategy would therefore be suitable for determining whether GJ~1132~b is a cloud-free exo-Venus.

For the longer observing time strategies, i.e. using 20 MIRI LRS eclipses, we find that $\gtrsim 3\sigma$ detections of both CO$_2$ and H$_2$O can be made across all three atmospheric compositions (Table \ref{tab:sig}). In particular, confident, $\sim4-8\sigma$ detections of these species can be made in the Venus-like and 50\%~CO$_2$/50\%~H$_2$O cases. The 100\%~H$_2$O case is slightly more challenging to characterise, allowing $\sim 3\sigma$ detections of H$_2$O with these observing strategies. We note, however, that the addition of four eclipses with the F1500W filter on top of the 20 MIRI LRS eclipses does not significantly improve the atmospheric constraints which are made. The spectra and temperature profiles are all retrieved to within 2$\sigma$, while the abundances of CO$_2$ and H$_2$O are accurately retrieved within 1$\sigma$ uncertainties of $\lesssim$0.75~dex. We therefore conclude that these longer observing time strategies would be able to characterise the atmosphere of GJ~1132~b across a range of CO$_2$- and H$_2$O-rich compositions, including determining the abundances of these two species. Given the importance of CO$_2$ and H$_2$O in geochemical processes, such constraints would be invaluable in determining the possible atmospheric origins of GJ~1132~b.

\subsection{Case Study: Trappist-1~b}

The Trappist-1 system \citep{Gillon2016,Gillon2017} is currently the most promising system of terrestrial-like exoplanets for atmospheric characterisation \citep{Turbet2020}. Thanks to the host star's small radius (0.117~R$_\odot$) and low effective temperature (2559~K), the secondary eclipse depths of the Trappist-1 planets are favourable despite the small planetary sizes and low planetary temperatures. In particular, Trappist-1~b is the warmest planet in this system, with an equilibrium temperature of 400~K, and represents an excellent candidate for atmospheric characterisation of a temperate exoplanet in thermal emission. Meanwhile, transmission spectroscopy observations have already begun to place constraints on the atmosphere of Trappist-1~b. \citet{deWit2016} rule out a clear, H$_2$-rich atmosphere at $>10\sigma$ and note that while a H$_2$-rich atmosphere with high-altitude clouds/hazes is allowed by the data, this is an unlikely scenario given the relatively low irradiation level of Trappist-1~b. Conversely, \citet{Bourrier2017} find a marginal decrease in the Lyman-$\alpha$ flux of Trappist-1 during the transits of planets b and c, which may indicate the presence of extended hydrogen exospheres. However, this effect could also be caused by stellar activity. A secondary atmosphere on Trappist-1b therefore remains a promising possibility, and we investigate the observability of a Venus-like, 50\%~CO$_2$/50\%~H$_2$O and 100\%~H$_2$O atmosphere.

Compared to LHS~3844~b and GJ~1132~b, Trappist-1~b's relatively small planetary radius and cooler temperature results in a lower S/N and the need for longer observation times. Here, we consider three observation strategies: (i) 30 eclipses with MIRI LRS, (ii) 80 eclipses with MIRI LRS, (iii) 80 eclipses with MIRI LRS plus 4 eclipses with the F1500W photometric band. The bottom panel of Figure \ref{fig:all_models} shows the simulated spectra for Trappist-1~b across the CO$_2$-rich to H$_2$O-rich compositions considered here, while the bottom three rows of Figure \ref{fig:ret_spectra} show the simulated data and retrieved spectra. For all three observational strategies, the data are clearly distinguishable from the bare-rock scenario with no energy redistribution. Furthermore, for the Venus-like composition, the sharp CO$_2$ feature at $\sim$9~$\mu$m can be distinguished by eye, consistent with the confident detections of CO$_2$ in these retrievals.

Assuming 30 MIRI LRS eclipses, we find that CO$_2$ and H$_2$O are detected in both the Venus-like and 50\%~CO$_2$/50\%~H$_2$O cases (Table \ref{tab:sig}). The Venus-like composition is most confidently constrained, with 5.5$\sigma$ and 4.9$\sigma$ detections of CO$_2$ and H$_2$O, respectively, while the 50\%~CO$_2$/50\%~H$_2$O composition leads to $\sim$3$\sigma$ detections of these species. However, in the 100\%~H$_2$O case the simulated data are consistent with a blackbody and spectral features are not detected with statistical significance. Nevertheless, for all three compositions the retrieval framework fits the true spectrum and photospheric $P$-$T$ profile within 2$\sigma$ uncertainties (Figure \ref{fig:ret_PT}). This 30-eclipse strategy would therefore be ideal to characterise (or rule out) clear, CO$_2$-dominated atmospheric compositions on Trappist-1~b.

Using a longer observing time of 80$-$84 eclipses, we find that the Venus-like and 50\%~CO$_2$/50\%~H$_2$O compositions are readily characterised, allowing $\sim$4-10$\sigma$ detections of CO$_2$ and H$_2$O (Table \ref{tab:sig}). The 100\%~H$_2$O composition is more challenging to characterise, though a blackbody spectrum is nevertheless rejected by the data at $\sim$4$\sigma$ with both of these observing strategies. For all three compositions, both the spectra and temperature profiles are retrieved within 2$\sigma$, while the abundances of CO$_2$ and H$_2$O are retrieved to within the 2$\sigma$ uncertainties. We note that these results are very similar for both the observing strategies with and without the photometry, suggesting that the MIRI LRS data is driving these detections. 80 eclipses using MIRI LRS are therefore sufficient to confidently characterise cloud-free CO$_2$-rich compositions for Trappist-1~b, and to detect the presence of atmospheric absorption in the case of a water-rich composition. Furthermore, across all of the observing strategies discussed here, the abundances of CO$_2$ and H$_2$O are constrained to within 1-$\sigma$ uncertainties of $<0.7$~dex, where these species are detected.

\begin{table*}
  \centering
  \caption{Detection significances for CO$_2$ and H$_2$O given different atmospheric compositions and observing strategies for LHS~3844~b, GJ~1132~b and Trappist-1~b (see Section \ref{sec:case_studies}). The confidence level at which the data eliminates a blackbody spectrum is also shown for each case (columns labelled `non-BB'). The number of eclipses with MIRI LRS and MIRI photometry assumed for each case are shown in italics. Confidence levels below 2$\sigma$ are not shown.}
    \begin{tabular}{lcccccccccc}
    \multicolumn{1}{c}{\multirow{2}[3]{*}{\textbf{Planet}}} & \multicolumn{3}{c}{\textbf{Venus}} &       & \multicolumn{3}{c}{\boldmath{}\textbf{50\% H$_2$O/50\% CO$_2$}\unboldmath{}} &       & \multicolumn{2}{c}{\boldmath{}\textbf{100\% H$_2$O}\unboldmath{}} \bigstrut[b]\\
\cline{2-4}\cline{6-8}\cline{10-11}          & \boldmath{}\textbf{CO$_2$}\unboldmath{} & \boldmath{}\textbf{H$_2$O}\unboldmath{} & \textbf{non-BB} &       & \boldmath{}\textbf{CO$_2$}\unboldmath{} & \boldmath{}\textbf{H$_2$O}\unboldmath{} & \textbf{non-BB} &       & \boldmath{}\textbf{H$_2$O}\unboldmath{} & \textbf{non-BB} \bigstrut\\
    \hline
    \hline
    \textbf{LHS 3844 b} &       &       &       &       &       &       &       &       &       &  \bigstrut[t]\\
    \emph{8 LRS} & 5.55  & 4.07  & 8.37  &       & $-$   & 2.52  & 2.28  &       & $-$   & 2.93 \\
    \emph{20 LRS} & 7.06  & 5.32  & 10.59 &       & 3.87  & 5.02  & 4.86  &       & 2.74  & 4.20 \\
    \emph{20 LRS, 4 F1500W} & 6.95  & 5.20  & 11.00 &       & 3.61  & 4.72  & 4.45  &       & 2.82  & 4.50 \bigstrut[b]\\
    \hline
    \textbf{GJ 1132 b} &       &       &       &       &       &       &       &       &       &  \bigstrut[t]\\
    \emph{8 LRS} & 4.70  & 4.42  & 5.69  &       & $-$   & 2.68  & 2.31  &       & $-$   & 4.38 \\
    \emph{20 LRS} & 7.62  & 6.91  & 8.75  &       & 4.35  & 4.53  & 5.03  &       & 3.18  & 3.20 \\
    \emph{20 LRS, 4 F1500W} & 8.05  & 6.90  & 9.20  &       & 4.76  & 5.05  & 5.33  &       & 2.96  & 3.20 \bigstrut[b]\\
    \hline
    \textbf{Trappist-1 b} &       &       &       &       &       &       &       &       &       &  \bigstrut[t]\\
    \emph{30 LRS} & 5.51  & 4.90  & 5.82  &       & 2.89  & 3.40  & 3.02  &       & 2.31  & 2.47 \\
    \emph{80 LRS} & 9.09  & 9.60  & 11.06 &       & 4.40  & 4.68  & 4.75  &       & 2.50  & 4.08 \\
    \emph{80 LRS, 4 F1500W} & 9.79  & 9.57  & 11.73 &       & 4.86  & 4.70  & 4.98  &       & 2.65  & 3.90 \bigstrut[b]\\
    \hline
    \end{tabular}%
  \label{tab:sig}%
\end{table*}%

\section{Discussion}
\label{sec:discussion}

In section \ref{sec:case_studies}, we have shown that rocky exoplanet atmospheres across a wide range of temperatures ($\sim$400-800~K) can be characterised in thermal emission with JWST/MIRI, including confident detections of atmospheric absorption by CO$_2$ and H$_2$O. Here, we begin by discussing the calculation of detection significances and key subtleties which can arise. Having focused on cloud- and haze-free atmospheric compositions in previous sections, we also discuss the impact which clouds and hazes may have on the characterisation of rocky exoplanet atmospheres in Section \ref{sec:discussion:clouds}. In sections \ref{sec:discussion:3D} and \ref{sec:discussion:stitching}, we also discuss 3D effects and stitching together multi-mode observations.

\subsection{Detection Significances}

When considering detection significances, it is useful to understand the role of model complexity in determining the Bayesian evidence. As described in Section \ref{sec:ret_detect}, the confidence of a molecular detection can be assessed by comparing the Bayesian evidences of two retrievals including/excluding the molecule(s) in question. Similarly, the confidence with which a blackbody spectrum can be rejected can be assessed by comparing the Bayesian evidences of retrievals with a full atmospheric model vs. only a single temperature (i.e. a blackbody model). However, these Bayesian evidences encapsulate not only the fit to the spectrum but also the complexity of the model. For example, if a 10-parameter model results in the same goodness of fit as a 5-parameter model, the 5-parameter model will have a higher Bayesian evidence as it has a lower model complexity. 

When calculating the detection significance of a single molecule, the models compared only have a difference of one parameter (i.e. the models are identical apart from the presence of one molecule), and so the detection significance calculated from this comparison is a fairly good measure of `goodness of fit' as the model complexities are comparable. However, when the `full model' is compared to a blackbody model, there is a significant difference in model complexity (i.e. 13 parameters for the full model vs 1 parameter for the blackbody). This means that the full model is penalised for its complexity relative to the blackbody model, and the blackbody spectrum may be rejected at a lower significance than expected. That is, the poor fit from a blackbody model may be compensated for by the simplicity of the model. 

An alternative way to assess how confidently a blackbody model can be rejected is to compare this model to a `core parameters' model. This `core parameters' model is based on the full atmospheric model, but includes only the species for which there is evidence in the data. For example, if H$_2$O and CO$_2$ are detected in the data (using the full model) but no other species are detected, the `core parameters' model would include only H$_2$O and CO$_2$ (as well as the usual temperature profile parameterisation). This effectively strips down the full model to the components which are necessary to fit the data, without including unnecessary parameters. Thus, when compared to the blackbody model, the `core parameters' model is not penalised by unnecessary parameters, and instead the difference in model complexity is more representative of the complexity required to fit the data. For example, in the Venus-like case for Trappist-1~b with 30 MIRI LRS eclipses, comparing the blackbody model to the `core parameters' model results in a 6.09$\sigma$ rejection of the blackbody, whereas a comparison to the full model results in a 5.82$\sigma$ rejection. Note that all detection significances shown in Table \ref{tab:sig} use a comparison with the full model.

Ultimately, it is important to understand how the models used can affect the confidence with which a blackbody spectrum can be rejected. In cases with very strong molecular detections (e.g. the LHS~3844~b case study shown here, first row of Table \ref{tab:sig}), a blackbody can be rejected with high confidence even when compared to the full model. In more marginal cases, however, it may be necessary to consider how model complexity affects this metric. We further note that confident molecular detections provide a robust way to reject a featureless spectrum as the model comparisons involved in this calculation are not subject to large differences in model complexity.

\subsection{Impact of Clouds and Hazes}
\label{sec:discussion:clouds}

Clouds and hazes can have significant effects on the temperature profiles and thermal emission spectra of low-mass exoplanets \citep[e.g.][]{Morley2015,Piette2020c}. So far in this work we have considered clear atmospheres, and we now discuss how clouds and hazes may affect the atmospheric characterisation of rocky exoplanets. While optical scattering from clouds and hazes is not directly visible in infrared observations, it has the effect of cooling the atmosphere and results in a more isothermal temperature profile. This can lead to a smaller planetary flux and muted spectral features, as demonstrated by \citet{Piette2020c} for mini-Neptunes. However, while cloudy atmospheres may provide challenges for chemical detections, secondary eclipse observations can still constrain atmospheric properties above the cloud deck, including the photospheric temperature \citep{Piette2020c}.

For the case studies LHS~3844~b, GJ~1132~b and Trappist-1~b, we consider the observability of the atmosphere given strong optical scattering from clouds and hazes. Figure \ref{fig:all_models} shows blackbody spectra for each of these planets at an equilibrium temperature which assumes a Venus-like Bond albedo of 0.7 and full day-night energy redistribution (light grey dashed lines). For LHS~3844~b, this spectrum would be detectable at $>$3$\sigma$ with MIRI LRS in 8 eclipses (e.g. for the resolution shown in Figure \ref{fig:ret_spectra}) or in the F1500W band in four eclipses. While spectral features may not necessarily be distinguishable due to the cloud opacity, the photospheric temperature could be constrained from such a detection. GJ~1132~b has a lower S/N ratio but would be detectable at $>$3$\sigma$ with 20 MIRI LRS eclipses (e.g. at the resolution shown in Figure \ref{fig:all_models}). Trappist-1~b would be more challenging to detect with MIRI LRS, but four eclipses in the F1500W band could be sufficient to detect the atmosphere and measure its photospheric temperature. 

We therefore conclude that the atmospheres of LHS~3844~b, GJ~1132~b and Trappist-1~b would be observable with MIRI even in the case of an extreme, Venus-like Bond albedo. Furthermore, a measurement of the photospheric temperature in this case would allow a joint constraint on the Bond albedo and day-night energy redistribution of the planet, which are degenerate in determining the dayside equilibrium temperature. For example, a photospheric temperature of $\sim$430~K for GJ~1132~b would suggest a high Bond albedo of $\sim$0.7/0.85, assuming efficient day-night energy redistribution/no day-night energy redistribution, respectively. In between the clear and Venus-like albedo cases, more moderate clouds and hazes may result in muted spectral features that can still be characterised using spectral retrieval analysis. JWST/MIRI observations of such rocky exoplanets will therefore place important constraints on their atmospheric properties, regardless of the presence of clouds and hazes.

\subsection{3-D effects}
\label{sec:discussion:3D}

3-dimensional effects are known to impact thermal emission observations of exoplanet atmospheres, especially those which are tidally-locked \citep[e.g.][]{Showman2009,Knutson2007,Demory2016,Kreidberg2019}. While efficient energy redistribution can act to homogenise the dayside atmosphere, dayside temperature and/or compositional inhomogeneities can have observable effects on the secondary eclipse spectrum of the atmosphere \citep[e.g.][]{Feng2020,Taylor2020,Cubillos2021}. For rocky exoplanets, a higher equilibrium temperature and/or lower surface pressure typically leads to less efficient day-night energy redistribution \citep{Koll2016}, with redistribution becoming efficient at surface pressures of $\mathcal{O}$(1)~bar \citep{Koll2019arXiv}. Therefore, the degree of longitudinal variation on the daysides of rocky exoplanets may vary considerably between different targets, depending on their atmospheric properties. In cases of more extreme inhomogeneity, it has been demonstrated for hot Jupiters that 1-D retrievals can result in biased results, depending on the signal-to-noise ratio of the data \citep[e.g.][]{Feng2020,Taylor2020,Cubillos2021}. It is therefore useful to place the results of 1-D atmospheric secondary eclipse retrievals into context, e.g. using multi-dimensional retrievals \citep[e.g.][]{Feng2016,Feng2020,Taylor2020} or with complementary phase curve observations. Nevertheless, 1-D retrievals can provide a valuable first glimpse into the dayside atmospheres of rocky exoplanets, especially considering the data quality expected for rocky exoplanets with JWST.

Ultimately, phase curve observations are required to deconstruct the longitudinal structure of an exoplanet atmosphere. Several methods exist to extract longitudinal information from such observations \citep[e.g.][]{Feng2016,Feng2020,Irwin2020,Cubillos2021}. In particular, \citet{Cubillos2021} present a new approach in which a 1-D retrieval is applied to each `slice' in a longitudinally-deconstructed phase curve, applying it to the phase curve of the hot Jupiter WASP-43~b. Such an approach could be used to extend \hydrar to interpret the phase curves of rocky exoplanets.

\subsection{Stitching Multi-Mode Observations}
\label{sec:discussion:stitching}

For each of the more time-intensive MIRI~LRS observing strategies considered in Section \ref{sec:case_studies}, we have investigated the impact of using MIRI LRS observations both with and without additional photometric observations with the F1500W filter. As discussed in Section \ref{sec:case_studies}, this additional photometric data does not significantly improve the chemical detections which can be made with MIRI~LRS alone. Additionally, it would be important to consider possible systematic offsets between different instrument modes, which are as yet uncharacterised for JWST and may lead to spurious detections (see e.g. \citealt{Barstow2015}). Future characterisation of the MIRI instrument may allow for the stitching of MIRI~LRS and MIRI photometry data, for example by calibrating these modes with multiple measurements of the same source at high S/N. Meanwhile, the cases explored here argue that a MIRI~LRS-only approach is able to characterise rocky exoplanet atmospheres with CO$_2$-rich to H$_2$O-rich compositions, independent of the photometry. We additionally note that, while effects due to stellar activity may hinder the stitching of transmission spectra, this is not the case for secondary eclipse spectra \citep[e.g.][]{Barstow2015}.


\section{Conclusions} 
\label{sec:conclusions}

In this work, we develop the first retrieval framework for thermal emission spectra of rocky exoplanets, \textsc{h}{\small y}\textsc{dr}{\small o}. This framework is able to retrieve the chemical composition of the atmosphere without assuming a particular background species, using the centered-log-ratio chemical abundance parameterisation of \citet{Benneke2012}. The method is therefore ideally suited for determining the chemistry of secondary atmospheres, with implications for the study of exo-geology. The retrieval framework is further able to constrain the atmospheric temperature profile, providing important information about the thermal properties and potential surface conditions of the planet.

In order to test the retrieval framework, we generate self-consistent atmospheric models and simulated JWST MIRI data for a range of irradiation conditions and atmospheric compositions. To do this, we adapt the \textsc{genesis} self-consistent atmospheric model \citep{Gandhi2017,Piette2020a,Piette2020c} for non-H$_2$ rich compositions. We first test the case of Trappist-1~b assuming a Venus-like composition and find that $\sim$30 secondary eclipses with MIRI LRS result in a confident CO$_2$ detection, consistent with previous studies \citep{Lustig-Yaeger2019}. The retrieval framework is able to accurately constrain the atmospheric chemical composition and temperature profile to within 2~$\sigma$, validating the method.

We then identify optimal rocky exoplanet candidates for atmospheric characterisation with JWST using a simple observability metric. The mid-infrared is an ideal spectral range for such efforts given the abundance of molecular features in this range and the relatively higher signal-to-noise ratio (S/N) in secondary eclipse. We therefore rank the observability of known rocky exoplanets by considering the number of eclipses required with MIRI LRS to detect the thermal emission spectrum at S/N=3 at a resolution of $R\sim$10, $E_{\rm S/N=3}$. This resolution is chosen to provide a metric for the spectroscopic observability of rocky exoplanets, which is relevant for the chemical characterisation of their atmospheres. As expected, host star brightness, planetary radius and planetary temperature play key roles in determining observability. While brighter host stars result in smaller observational uncertainties, larger planetary radii and temperatures result in a greater signal for a given stellar radius and temperature.

We find that $>30$ known rocky exoplanets should have detectable thermal emission signatures with MIRI LRS at S/N=3 and a resolution of $R\sim$10. Of these, 14 are more observable than GJ~1132~b, which is considered to be an optimal target for rocky exoplanet atmospheric characterisation \citep{Morley2017b}. In particular, we find that LHS~3844~b, GJ~1132~b and Trappist-1~b each represent ideal candidates for the hot ($\sim$1000~K), warm ($\sim$600~K) and temperate ($\sim$300-400~K) regimes, respectively. We therefore focus on these case studies to assess the observability of secondary atmospheres across this range of irradiation conditions.

We investigate the observational strategies required to characterise a range of CO$_2$-rich to H$_2$O-rich atmospheres in LHS~3844~b, GJ~1132~b and Trappist-1~b. In particular, we consider four key science cases in order of scientific complexity which may be addressed by thermal emission observations of these atmospheres: (i) Is the spectrum consistent with a hot, bare rock surface (assuming no heat redistribution, as investigated by \citealt{Koll2019b})? (ii) Are spectral features detected in the spectrum, confirming atmospheric absorption? i.e. Can a blackbody be ruled out by the data? (iii) Is CO$_2$ confidently detected in the spectrum, e.g. using the prominent spectral feature at $\sim$9~$\mu$m? (iv) Is H$_2$O confidently detected?

We investigate the confidence with which these science cases can be addressed for a cloud-free Venus-like (CO$_2$-rich), 50\% CO$_2$/50\% H$_2$O or 100\% H$_2$O atmosphere for the case studies LHS~3844~b, GJ~1132~b and Trappist-1~b. For LHS~3844~b, we find that 8 MIRI LRS eclipses are sufficient to confidently characterise a cloud-free Venus-like composition, leading to $>4\sigma$ detections of CO$_2$ and H$_2$O. Alternatively, 20 eclipses with MIRI~LRS allow both a Venus-like and 50\% CO$_2$/50\% H$_2$O composition to be confidently characterised, while a blackbody spectrum would be significantly rejected (at $>4\sigma$) in the case of a 100\%~H$_2$O composition, indicating atmospheric absorption. Furthermore, in the cases where CO$_2$ and H$_2$O are detected, their abundances can be constrained within $\sim 0.5$~dex, i.e. comparable to the precision currently achieved for hot Jupiters \citep[e.g.][]{Sheppard2017,Welbanks2019b}.

For GJ~1132~b, we find that 8 eclipses with MIRI LRS allow $>4 \sigma$ detections of CO$_2$ and H$_2$O for the Venus-like composition, and can rule out a blackbody spectrum at $\sim 4\sigma$ for a 100\%~H$_2$O composition. Meanwhile, a longer observing time strategy of 20 MIRI~LRS eclipses allows $>3\sigma$ detections of CO$_2$ and H$_2$O across the CO$_2$- to H$_2$O-dominated compositions considered here. Furthermore, in the cases where CO$_2$ and H$_2$O are detected, their abundances are constrained to within 1$\sigma$ uncertainties of $\lesssim$0.75~dex.

In the case of Trappist-1~b, we find that a Venus-like composition could be readily characterised with 30 MIRI~LRS eclipses, consistent with previous studies \citep{Lustig-Yaeger2019}. In particular, we find that $\sim 5\sigma$ detections of CO$_2$ and H$_2$O would be possible for a Venus-like composition and that, in the case of a 50\% CO$_2$/50\% H$_2$O composition, H$_2$O could be detected at $\sim$3$\sigma$. We also consider a longer observing time strategy of 80 eclipses with MIRI~LRS, and find that this leads to $>4\sigma$ detections of CO$_2$ and H$_2$O across the Venus-like and 50\% CO$_2$/50\% H$_2$O compositions, while a blackbody spectrum is rejected at $>4\sigma$ in the case of a 100\%~H$_2$O composition. For both of these observing strategies, the abundances of CO$_2$ and H$_2$O (in the cases where they are detected) are constrained to within a 1-$\sigma$ precision of $<0.7$~dex. In addition to the chemical constraints described above, we find that the photospheric temperature profiles of LHS~3844~b, GJ~1132~b and Trappist-1~b are all accurately retrieved within the 2$\sigma$ uncertainties, with temperature constraints as tight as $\sim$100~K in some cases. The observing strategies described above for each of these planets may therefore provide an unprecedented view into the thermal conditions on these rocky worlds.  

We also find that for the longer observing time strategies investigated here, an additional 4 eclipses with the F1500W photometry band do not contribute significantly to the constraints achieved with the MIRI LRS data alone. The degeneracies between opacity from different molecules such as CO$_2$ and H$_2$O make it challenging to place chemical constraints on the atmosphere using photometry. For example, while the F1500W bandpass probes a strong CO$_2$ feature, it is also impacted by a broad, continuum feature of H$_2$O. Therefore, a MIRI~LRS-only observing strategy may be ideal for the initial chemical characterisation of rocky exoplanets. However, mid-infrared photometric observations are valuable as quick tests for the presence of an atmosphere as they are able to distinguish scenarios with/without atmospheric energy redistribution (\citealt{Koll2019b}, see also Figure \ref{fig:all_models}).

The temperature profiles we explore in this work are for clear atmospheres, though we note that clouds and hazes may affect the temperature profile and resulting thermal emission spectrum. For example, optical scattering from clouds and hazes can result in somewhat muted spectral features \citep[e.g.][]{Morley2015,Piette2020c}. For the case studies LHS~3844~b, GJ~1132~b and Trappist-1~b, we find that photospheric temperatures could nevertheless be measured with MIRI~LRS and/or MIRI photometry in the case of a high, Venus-like Bond albedo of 0.7. Such a measurement would allow joint constraints on the Bond albedo and energy redistribution efficiency in the atmosphere, and very cool photospheric temperatures (i.e. high Bond albedos) may suggest the presence of clouds. Thermal emission observations will therefore provide a unique opportunity to probe the conditions in cloudy, rocky exoplanet atmospheres.

The characterisation of rocky exoplanet atmospheres arguably represents the next frontier in exoplanet science. We have shown here that secondary eclipse observations with JWST/MIRI will be able to place important constraints on atmospheric chemistry of rocky exoplanets and potential geochemical signatures. Furthermore, we find that $>$30 known rocky exoplanets are suitable candidates for dayside atmospheric characterisation with JWST/MIRI. Thermal emission spectroscopy offers an unparalleled opportunity to characterise secondary atmospheres, whose small scale heights can be prohibitive in transmission spectroscopy, and will be an essential tool to characterise exo-geology in the near future.

\section*{Acknowledgements}

We thank the anonymous reviewer for their thorough and helpful comments on our manuscript. We also thank Subhajit Sarkar for help with validating our PandExo uncertainties against JexoSim. AAAP acknowledges support from the UK Science and Technology Facilities Council (STFC) towards her doctoral studies. This research has made use of the NASA Exoplanet Archive, which is operated by the California Institute of Technology, under contract with the National Aeronautics and Space Administration under the Exoplanet Exploration Program.

\section*{Data availability}
No new data were generated or analysed in support of this research.




\bibliographystyle{mnras}
\bibliography{main} 






\appendix

\section{Retrieval with higher surface pressure}
\label{sec:highP_appendix}

\begin{figure*}
    \centering
    \includegraphics[width=0.45\textwidth]{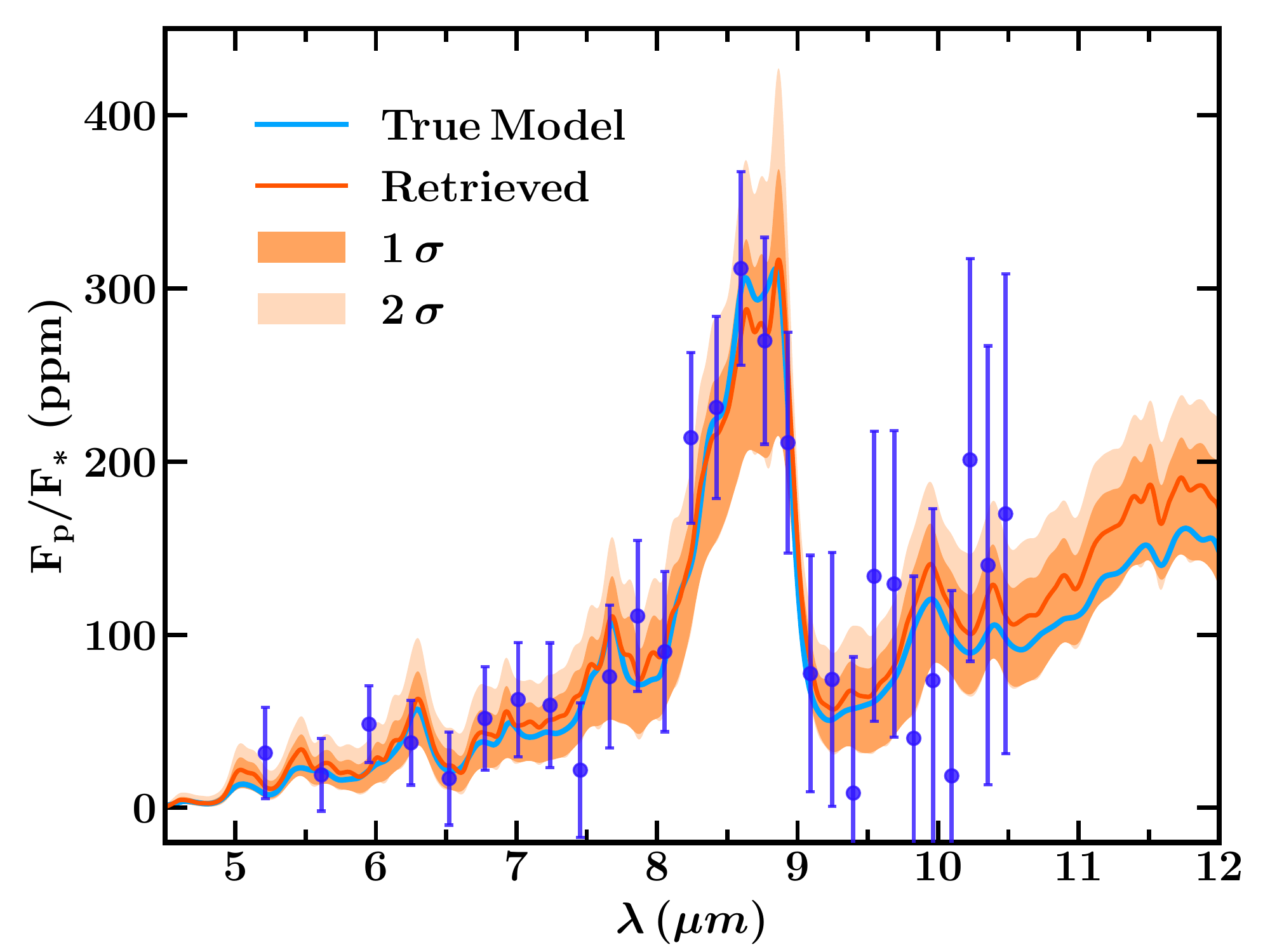}
    \includegraphics[width=0.45\textwidth]{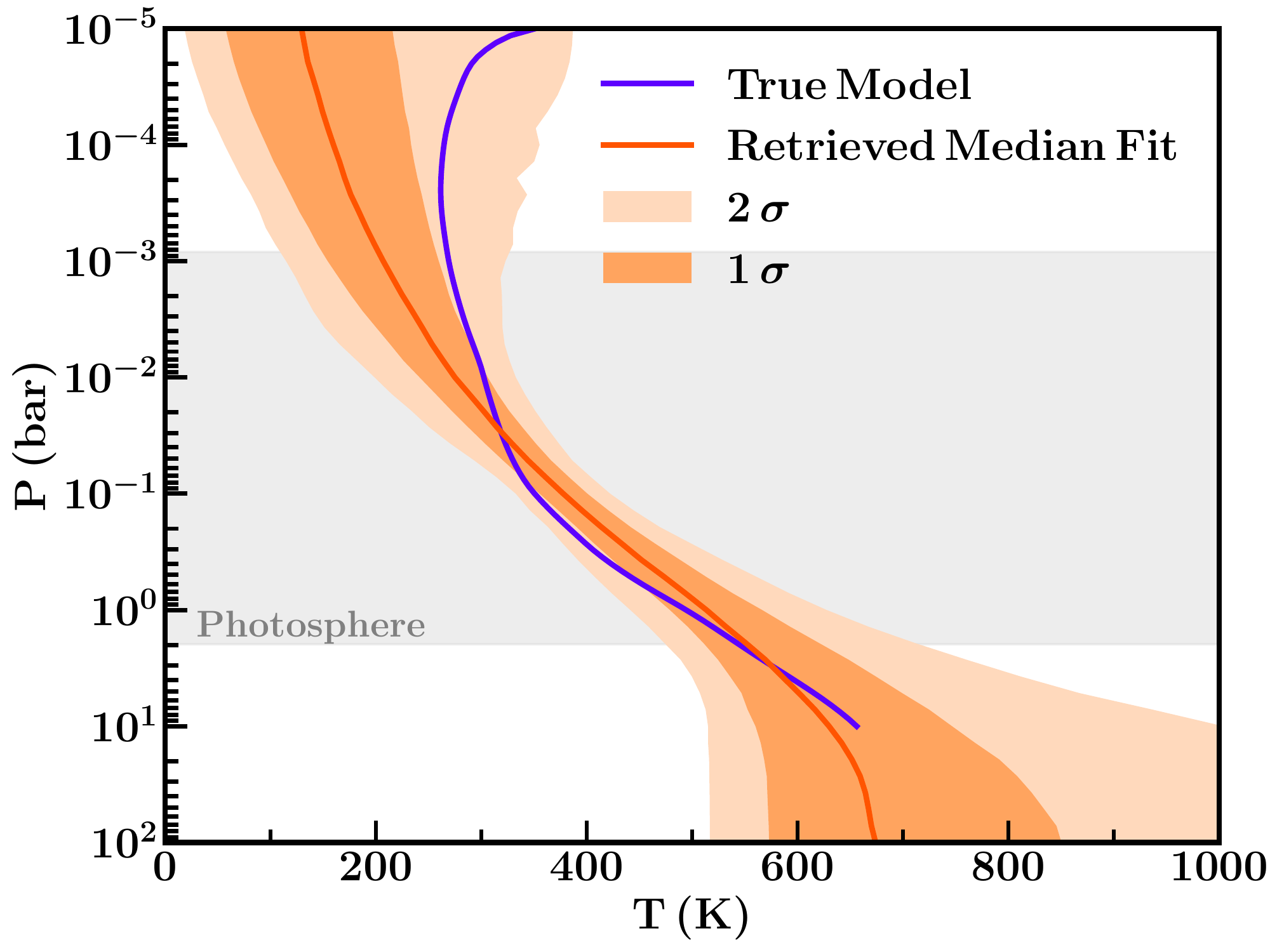}
    \includegraphics[width=0.95\textwidth]{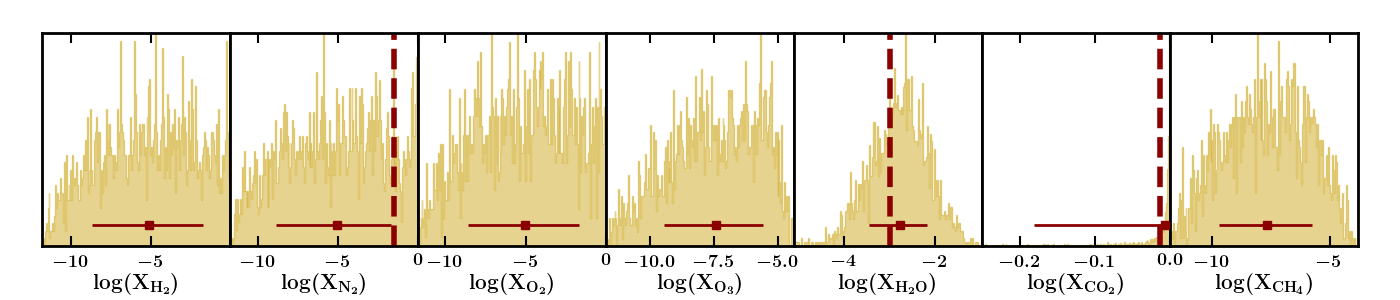}
    \caption{Same as Figure \ref{fig:validation} but assuming a surface pressure of 100~bar for the retrieval model, rather than 10~bar. The simulated data is identical to that used in Figure \ref{fig:validation}. The retrieved atmospheric properties are not sensitive to the difference in surface pressure used in the retrieval model.}
    \label{fig:validation_highP}
\end{figure*}

Figure \ref{fig:validation_highP} shows the results of a retrieval for Trappist-1~b similar to that in Section \ref{sec:validation}, but assuming a surface pressure of 100~bar rather than 10~bar in the retrieval model. Both of these surface pressures are deeper than the photosphere and the resulting spectra are therefore not sensitive to the surface pressure. This is demonstrated by the fact that the retrieved atmospheric properties in Figure \ref{fig:validation_highP} are the same as those in Figure \ref{fig:validation}, despite the difference in the surface pressure assumed for the retrieval model.

\section{Marginalised posterior probability distributions for molecular abundances}
\label{sec:corner_plot_appendix}

\begin{figure*}
    \centering
    \includegraphics[width=0.8\textwidth]{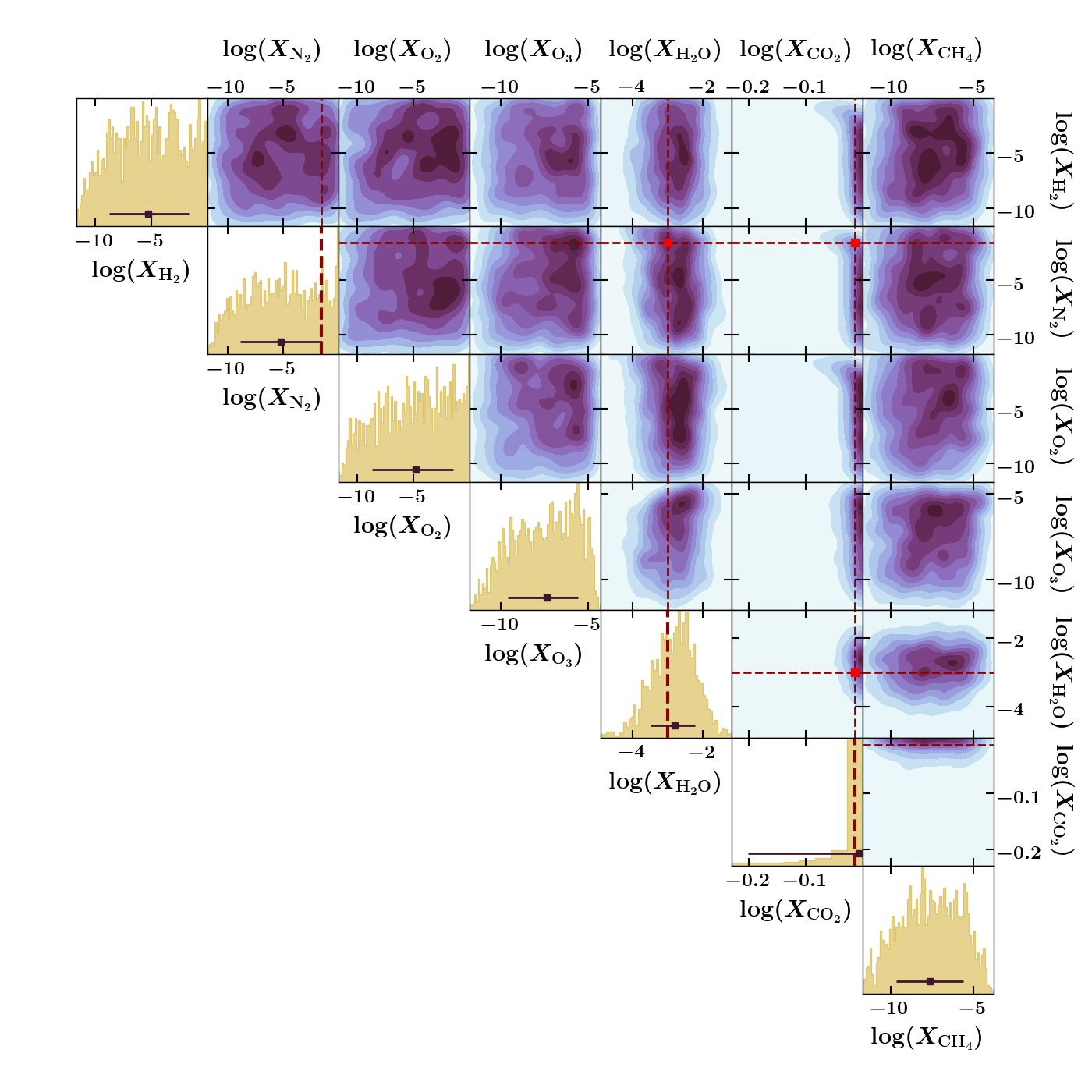}
    \caption{1D and 2D posterior probability distributions corresponding to the validation retrieval shown in Section \ref{sec:validation} and Figure \ref{fig:validation}. The diagonal shows the 1D marginalised posterior probability distributions for each molecular abundance, while off-diagonal panels show 2D marginalised posterior probability distributions. Red squares and dashed lines indicate the true input abundances of CO$_2$, N$_2$ and H$_2$O. Dark brown squares and error bars show the median and 1-sigma uncertainties in the retrieved abundances.}
    \label{fig:corner_plot}
\end{figure*}

Figure \ref{fig:corner_plot} shows the 1D and 2D marginalised posterior probability distributions of the retrieved chemical abundances for the validation retrieval of Trappist-1~b discussed in Section \ref{sec:validation}.

\section{Optimal targets for rocky exoplanet thermal emission}
\label{sec:best_targets_appendix}
In Table \ref{tab:best_targets}, we list the known rocky exoplanets whose thermal emission in secondary eclipse can be observed to S/N=3 in fewer than 10 eclipses at $R\sim$10. We have explored the potential for atmospheric characterisation of LHS~3844~b and GJ~1132~b in Section \ref{sec:case_studies} and found that as few as 8 MIRI~LRS eclipses are sufficient to detect species such as CO$_2$ and H$_2$O depending on the composition of the atmosphere. This list of promising targets therefore represents a population of rocky exoplanets whose atmospheres may be readily characterised in thermal emission using JWST MIRI.

\begin{table*}[htbp]
  \centering
  \caption{Properties of known rocky exoplanets whose thermal emission in secondary eclipse can be measured to S/N=3 (at $R\sim$10) in $\leq$10 eclipses with MIRI LRS (see Section \ref{sec:targets}), listed in order of observability. Note that we do not include 55~Cnc~e in this list as it would saturate part of the MIRI LRS detector.}
    \begin{tabular}{cccccccc}
    \hline
    \textbf{Planet} & \boldmath{}\textbf{$R_\mathrm{p}$ ($R_\oplus$)}\unboldmath{} & \boldmath{}\textbf{$T_\mathrm{eq}$ (K)}\unboldmath{} & \boldmath{}\textbf{$R_\mathrm{s}$ ($R_\odot$)}\unboldmath{} & \boldmath{}\textbf{$T_\mathrm{eff}$ (K)}\unboldmath{} & \textbf{K mag} & \textbf{Eclipses for S/N=3} & \textbf{Refs} \bigstrut\\
    \hline
    \hline
    LHS 3844 b & 1.30  & 809   & 0.19  & 3036  & 9.14  & 0.16  & 1,2 \bigstrut[t]\\
    HD 3167 b & 1.70  & 1752  & 0.86  & 5261  & 7.07  & 0.17  & 1,3 \\
    HD 219134 b & 1.60  & 1020  & 0.78  & 4699  & 3.26  & 0.17  & 1,4 \\
    K2-141 b & 1.54  & 2046  & 0.67  & 4373  & 8.40  & 0.19  & 1,5 \\
    GJ 9827 b & 1.58  & 1186  & 0.60  & 4340  & 7.19  & 0.26  & 1,6 \\
    HD 213885 b & 1.75  & 2138  & 1.10  & 5978  & 6.42  & 0.28  & 1,7 \\
    GJ 1252 b & 1.19  & 1091  & 0.39  & 3458  & 7.92  & 0.29  & 1,8 \\
    LTT 3780 b & 1.33  & 891   & 0.37  & 3331  & 8.20  & 0.48  & 1,9 \\
    HD 219134 c & 1.51  & 786   & 0.78  & 4699  & 3.26  & 0.55  & 1,4 \\
    HD 80653 b & 1.61  & 2421  & 1.22  & 5838  & 8.02  & 0.59  & 1,10,11 \\
    Kepler-21 b & 1.64  & 2034  & 1.90  & 6305  & 6.95  & 0.73  & 1, 12 \\
    L 168-9 b & 1.39  & 985   & 0.60  & 3800  & 7.08  & 0.97  & 1,13 \\
    HD 15337 b & 1.70  & 993   & 0.84  & 5131  & 7.04  & 1.23  & 1,14 \\
    GJ 1132 b & 1.13  & 586   & 0.21  & 3270  & 8.32  & 1.32  & 1,15 \\
    GJ 357 b & 1.22  & 528   & 0.34  & 3505  & 6.47  & 1.64  & 1,16 \\
    K2-291 b & 1.59  & 1403  & 0.90  & 5520  & 8.35  & 1.85  & 1,17 \\
    L 98-59 c & 1.35  & 516   & 0.31  & 3412  & 7.10  & 2.20  & 1,18 \\
    TOI-1235 b & 1.74  & 758   & 0.63  & 3872  & 7.89  & 2.47  & 1,19 \\
    LTT 1445 A b & 1.38  & 438   & 0.28  & 3337  & 6.50  & 2.59  & 1,20 \\
    Kepler-10 b & 1.47  & 2190  & 1.06  & 5708  & 9.50  & 2.74  & 1,21 \\
    WASP-47 e & 1.82  & 2209  & 1.16  & 5576  & 10.19 & 3.25  & 1,22 \\
    GJ 9827 c & 1.24  & 821   & 0.60  & 4340  & 7.19  & 3.54  & 1,6 \\
    CoRoT-7 b & 1.58  & 1766  & 0.82  & 5259  & 9.81  & 3.85  & 1,23,24 \\
    K2-265 b & 1.71  & 1427  & 0.98  & 5477  & 9.26  & 3.87  & 1,25 \\
    K2-229 b & 1.16  & 1964  & 0.79  & 5185  & 9.05  & 3.94  & 1,26 \\
    Kepler-93 b & 1.48  & 1140  & 0.92  & 5655  & 8.37  & 4.13  & 1,27,28 \\
    EPIC 220674823 b & 1.52  & 2291  & 0.87  & 5470  & 10.34 & 4.14  & 1,29 \\
    K2-216 b & 1.75  & 1104  & 0.72  & 4503  & 9.72  & 4.43  & 1,30 \\
    K2-111 b & 1.90  & 1233  & 1.23  & 5730  & 9.38  & 5.69  & 1,31 \\
    K2-36 b & 1.43  & 1351  & 0.72  & 4916  & 9.45  & 5.79  & 1,32 \\
    LHS 1140 c & 1.28  & 436   & 0.21  & 3216  & 8.82  & 9.73  & 1,33 \\
    L 98-59 d & 1.57  & 408   & 0.31  & 3412  & 7.10  & 9.84  & 1,18 \bigstrut[b]\\
    \hline
    \end{tabular}%
    \begin{tablenotes}
    \item \footnotesize Values obtained from: 1. \citet{Skrutskie2006}, 2. \citet{Vanderspek2019}, 3. \citet{Christiansen2017}, 4. \citet{Gillon2017b}, 5. \citet{Barragan2018}, 6. \citet{Rice2019}, 7. \citet{Espinoza2020}, 8. \citet{Shporer2020}, 9. \citet{Cloutier2020}, 10. \citet{Frustagli2020}, 11. \citet{Gaia2018}, 12. \citet{Lopez-Morales2016}, 13. \citet{Astudillo-Defru2020}, 14. \citet{Dumusque2019}, 15. \citet{Bonfils2018}, 16. \citet{Luque2019}, 17. \citet{Kosiarek2019}, 18. \citet{Cloutier2019b}, 19. \citet{Cloutier2020b}, 20. \citet{Winters2019}, 21. \citet{Weiss2016}, 22. \citet{Almenara2016}, 23. \citet{Barros2014}, 24. \citet{Dai2019}, 25. \citet{Lam2018}, 26. \citet{Santerne2018}, 27. \citet{Dressing2015}, 28. \citet{Stassun2019}, 29. \citet{Guenther2017}, 30. \citet{Persson2018}, 31. \citet{Fridlund2017}, 32. \citet{Damasso2019}, 33. \citet{Ment2019}.
    \end{tablenotes}
  \label{tab:best_targets}%
\end{table*}%


\bsp	
\label{lastpage}
\end{document}